\documentclass[%
 reprint,
superscriptaddress,
showpacs,
showkeys,
 amsmath,amssymb,
 aps,
 pra
]{revtex4-1}

\usepackage{graphicx}%
\usepackage{dcolumn}%
\usepackage{bm}%
\usepackage{subfig}%
\usepackage{floatrow} %
\usepackage{placeins}
\usepackage{romannum}
\usepackage{gensymb}

\newcommand{\SM}[1]{#1}
\newcommand{\FIG}[1]{Fig.~\ref{#1}}

\begin{document}

\title{The collective effect of finite-sized inhomogeneities on the spatial spread of populations in two dimensions}%

\author{Wolfram M\"obius}
\thanks{These two authors contributed equally. WM's current affiliations are Living Systems Institute and Physics and Astronomy at University of Exeter.}
\affiliation{Living Systems Institute, University of Exeter, Exeter, United Kingdom}
\affiliation{Physics and Astronomy, College of Engineering, Mathematics and Physical Sciences, University of Exeter, Exeter, United Kingdom}
\affiliation{Department of Applied Physics, Technische Universiteit Eindhoven, Eindhoven, The Netherlands}
\affiliation{Department of Physics, Harvard University, Cambridge, MA, USA}
\author{Francesca Tesser}
\thanks{These two authors contributed equally. WM's current affiliations are Living Systems Institute and Physics and Astronomy at University of Exeter.}
\affiliation{Department of Applied Physics, Technische Universiteit Eindhoven, Eindhoven, The Netherlands}
\affiliation{PMMH, ESPCI Paris - PSL, Paris, France}
\author{Kim M. J. Alards}
\affiliation{Department of Applied Physics, Technische Universiteit Eindhoven, Eindhoven, The Netherlands}
\author{Roberto Benzi}
\affiliation{Universit\'a di Roma ``Tor Vergata" and INFN, Rome, Italy}
\author{David R. Nelson}
\affiliation{Department of Physics, Harvard University, Cambridge, MA, USA}
\affiliation{Department of Molecular and Cellular Biology, Harvard University, Cambridge, MA, USA}
\author{Federico Toschi}
\affiliation{Department of Applied Physics, Technische Universiteit Eindhoven, Eindhoven, The Netherlands}
\affiliation{Istituto per le Applicazioni del Calcolo, Consiglio Nazionale delle Ricerche, Rome, Italy}

\date{\today}%

\begin{abstract} %
The dynamics of a population expanding into unoccupied habitat has been primarily studied for situations in which growth and dispersal parameters are uniform in space or vary in one dimension. Here we study the influence of finite-sized individual inhomogeneities and their collective effect on front speed if randomly placed in a two-dimensional habitat. We use an individual-based model to investigate the front dynamics for a region in which dispersal or growth of individuals is reduced to zero (obstacles) or increased above the background (hotspots), respectively. In a regime where front dynamics is determined by a local front speed only, a principle of least time can be employed to predict front speed and shape. The resulting analytical solutions motivate an event-based algorithm illustrating the effects of several obstacles or hotspots. We finally apply the principle of least time to large heterogeneous environments by solving the Eikonal equation numerically. Obstacles lead to a slow-down that is dominated by the number density and width of obstacles, but not by their precise shape. Hotspots result in a speedup, which we characterise as function of hotspot strength and density. Our findings emphasise the importance of taking the dimensionality of the environment into account.

\end{abstract}


\maketitle

Populations spread into yet-unoccupied habitats on a wide range of length and time scales. Prominent examples are the spread of invasive plants on large spatial scales and the growth of microbial populations on small spatial scales. Despite being so different at first sight, all these population expansions are driven by two processes, population growth and active or passive dispersal \cite{ShigesadaBook,LewisBook,LutscherBook}. While the former drives overall growth of the population, i.e., the number of individuals, the latter is necessary for the population to spread into new habitat.

\begin{figure}
\includegraphics[width=\linewidth]{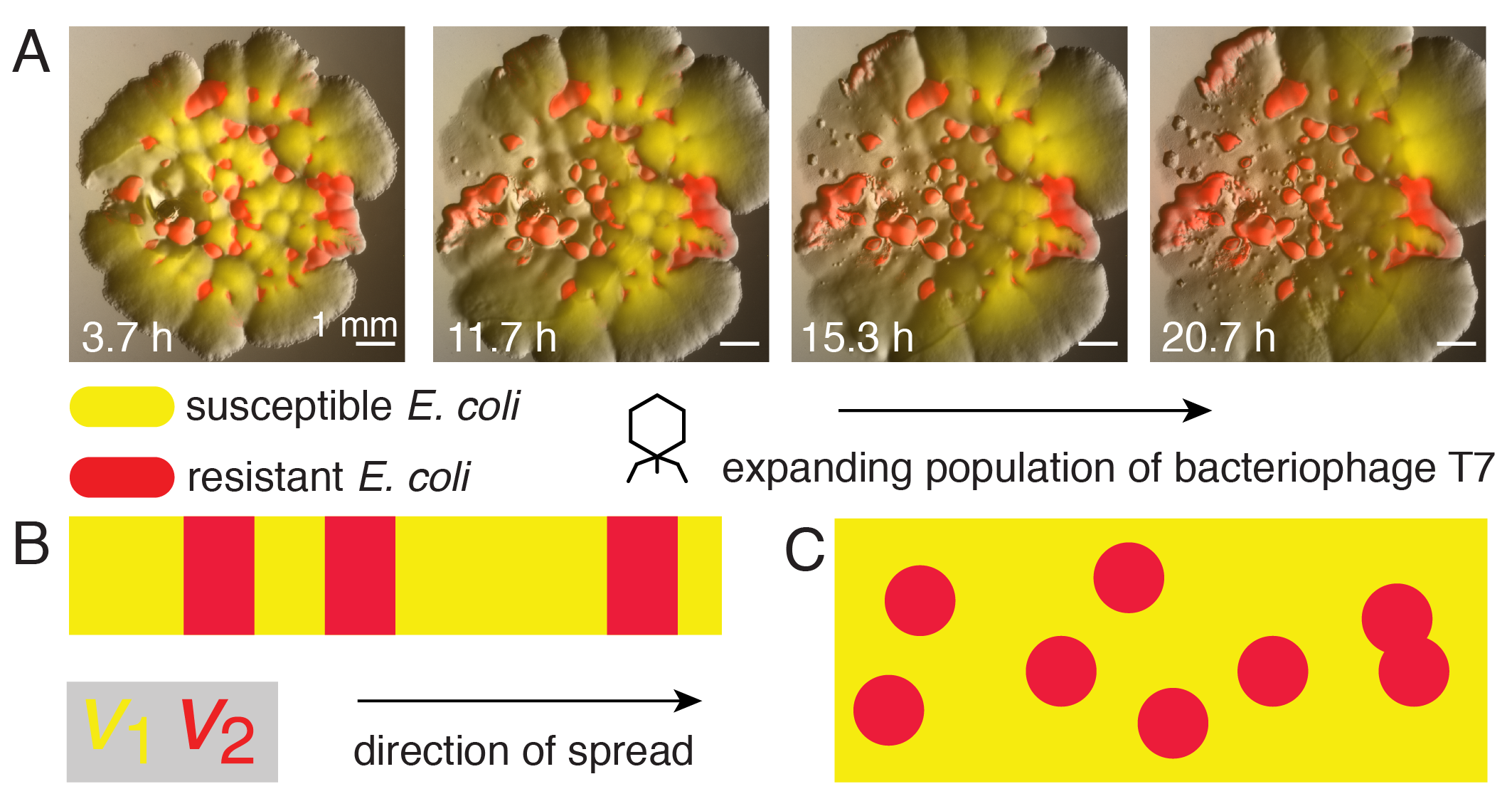}
\caption{\textbf{(A)} Experimental realisation of a population front encountering a heterogeneous environment. A population of bacteriophage T7 (dark area) is expanding on a lawn of \textit{E. coli}, where yellow areas represent patches of bacteria which can be infected by the bacteriophage (i.e., in which the population front can expand) while red areas represent patches of \textit{E. coli} which are known to be resistant (see Ref. \cite{MoebiusPLOSCB15} for a description of the experiment and additional information). \textbf{(B)} Sketch of an effectively one-dimensional heterogeneous environment where red and yellow patches differ in their support for population expansion by allowing different expansions speeds, $v_1$ and $v_2$, respectively. \textbf{(C)} Like panel (B), but for a two-dimensional environment.}
\label{fig:fig1}
\end{figure}

The environment encountered by these populations is often heterogeneous, i.e., the growth or dispersal processes may vary locally. An example is displayed in \FIG{fig:fig1}A: A population of a bacterial virus is expanding in a heterogeneous environment consisting of two types of bacteria. A region of bacteria which supports growth of the virus population (indicated in yellow, by use of yellow fluorescent proteins inside bacteria) is interspersed with regions of bacteria  that do not support growth of the phage population (indicated in red) \cite{MoebiusPLOSCB15}.

Much work has focused on heterogeneous one-dimensional environments such as depicted in \FIG{fig:fig1}B, where yellow and red indicate two different kinds of patches with specified population growth and dispersal, see, e.g., Ref. \cite{LutscherTPB20,RamanantoaninaMB16,AzimzadeSciRep20} and therein. For example, considering linear periodic habitats, Shigesada et al.~\cite{ShigesadaTPB86} studied invasion conditions of migrating species and the resulting periodic travelling waves. Limiting oneself to one-dimensional space not only simplifies the theoretical treatment, but also describes expansions in linear habitats such as along coastlines, watercourses or transportation networks.

Care has to be taken when generalising the results from studies of one-dimensional environments to higher dimensions. This is because results from linear habitats generally cannot be easily transferred: Consider \FIG{fig:fig1}B with a scenario where the red patches slow down an invasion so it almost comes to a halt. Due to the alternating position of red and yellow patches, these isolated red patches thus have a dramatic influence on the overall invasion process. The situation is different in two dimensions if the red patches are of finite size, yet isolated, as in \FIG{fig:fig1}C. In this case, as we will show, they affect the overall invasion process only marginally for low to intermediate densities, because invading populations can envelope finite-sized obstacles. Two-dimensional habitats are realised at the surfaces of solid substrates or liquids. Accordingly, our findings may find applications in the field of landscape ecology of invasive spread \cite{WithCB02}, complementing existing simulation-based work \cite{McInernyEcolInfo07,RodriguesMMNP13,CollinghamEcolAppl00,BocediBiorxiv20}. In addition, effectively two-dimensional populations can be found embedded in other environments, such as thin phytoplankton layers in the ocean \cite{DurhamAnnuRevMS12}. 

We here consider two different types of inhomogeneities. They may be associated with a population growth rate that is different to that of the embedding environment or may be regions within which dispersal of individuals differs. We find that significant progress can be made in a regime where the locally varying growth and dispersal properties result in a well-defined locally varying front speed that is independent of front speed at other locations or times. This regime has an analogy in geometrical optics where the refractive index and thus the speed of light vary locally. In consequence, our findings for front propagation in the presence of finite-sized inhomogeneities may be relevant for a range of propagation phenomena that share the trait of locally varying front speed, but not necessarily the underlying mechanism for front propagation: the spread of bacteriophage on a bacterial lawn (\FIG{fig:fig1}A), invasive brain tumours for which it is essential to differentiate tumour cell motility in white and grey matter \cite{HarpoldJNEN07}, the propagation of flame fronts \cite{ProvatasPRE95}, and autocatalytic reactions in porous media \cite{AtisPRL13}.

Note that locally varying dispersal does not necessarily mean that individuals move differently. Under certain circumstances, e.g., slow reaction or small-scale turbulence (thickness of the front much broader than the scale of turbulent eddies), the effect of turbulent background flows can also be described by an effective total diffusivity \cite{BrandenburgPRE11}. Thus, the example of turbulent patches with a position-dependent effective diffusivity broadens the scenarios we consider.

Our findings build on recent studies that considered isolated obstacles to two-dimensional population expansions \cite{MoebiusPLOSCB15} and expansions over curved surfaces \cite{BellerEPL18}, but expands beyond them: We here consider consider the converse of obstacles to invasions and characterise their consequences. Furthermore, instead of focusing on individual inhomogeneities, we investigate a whole range of environments, from those with isolated inhomogeneities to environments where features are so abundant that they almost fill up the two-dimensional space. The features considered are of finite size and randomly distributed, complementary to work focusing on purely random, two-dimensional periodic, and fractal-based environments \cite{GralkaELife19,KinezakiTPB10,HodgsonPLOSONE12}.

\section*{Individual-based simulations}
\label{sec:pbmodel}

An expanding population can be described at different levels of detail or coarsening. We first consider an individual-based scheme which allows us to take discreteness and random fluctuations into account directly. Individuals in the population can undergo growth and dispersal~\cite{PigolottiTPB13}, whereby the growth process includes both birth and death of individuals.

\begin{figure*}
\includegraphics{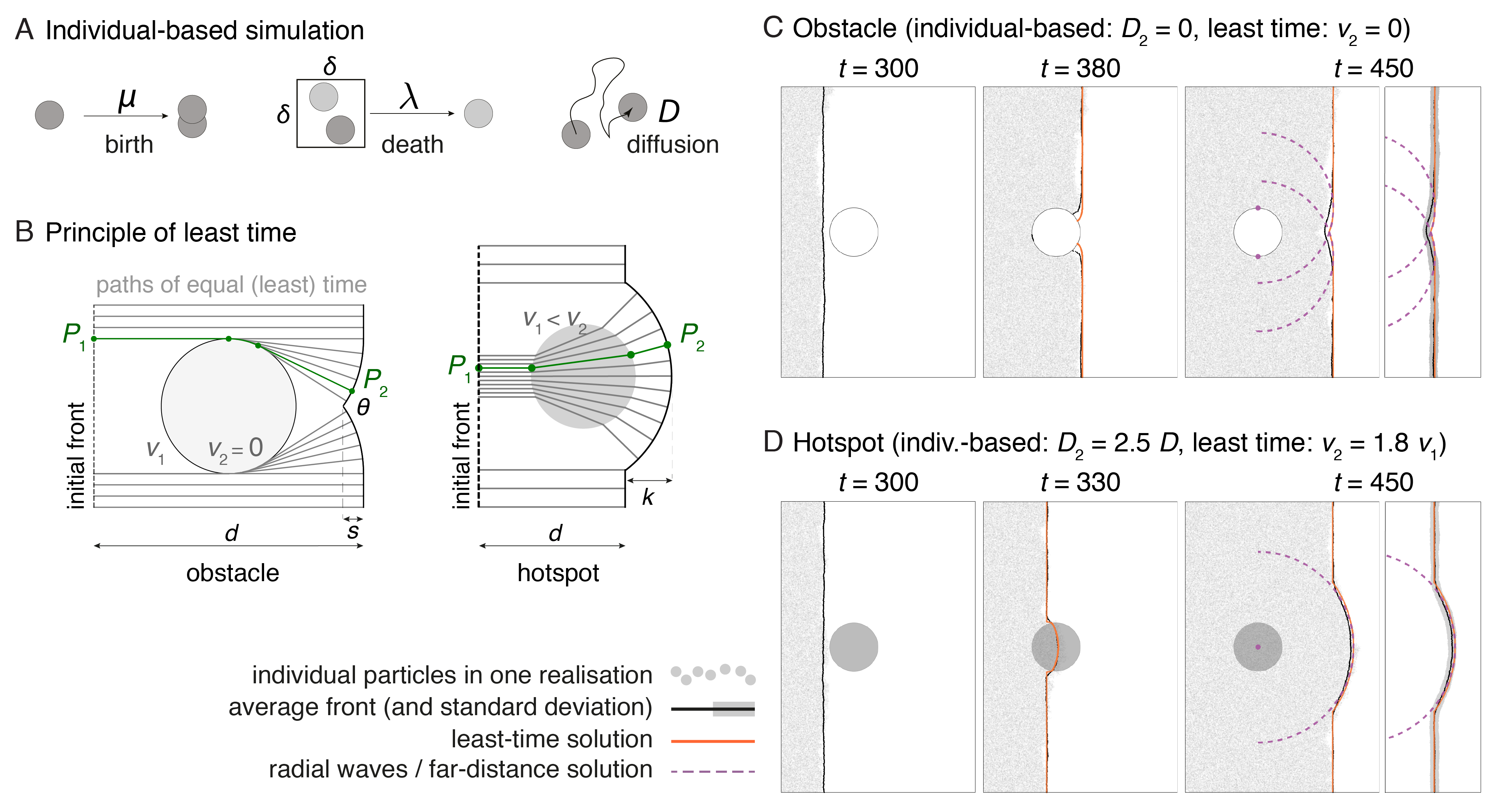}
\caption{\textbf{(A)} Sketch of the individual-based model with birth, death, and diffusion. Duplication of individuals occurs with rate $\mu$, death by competition within a squared cell of size $\delta^2$ with two individuals at rate $\lambda$. $D$ is the diffusion coefficient. \textbf{(B)} Least-time consideration for an obstacle (left, Ref. \cite{MoebiusPLOSCB15}) and a hotspot (right, \SM{Appendix S2}). There, the green line is one example for a path of least time from point $P_2$ back to the initial condition, which is reached at point $P_1$. The other grey lines represent paths of virtual markers traveling from left to right in the same amount of time. \textbf{(C)} Results of the individual-based simulation with an obstacle (white circle) with radius $R = 50$ and $D_2=0$ (grey dots), overlaid by the average front obtained from multiple realisations (black line, outside the obstacle), the least-time solution (orange line), and the far-distance solution (radial waves, purple dashed lines), see also (\SM{Video~S1}, Ref.~\cite{MoebiusPLOSCB15}). Right-most panel indicates standard deviation to average front instead of individual particles. \textbf{(D)} Similar to panel \textbf{(C)}, but the obstacle is replaced by a hotspot (grey circle) with radius $R=50$ and $D_2=2.5 D$ (\SM{Video S2}).}
\label{fig:fig2}
\end{figure*}

Birth is a duplication of an existing individual without change of position that occurs at rate $\mu$. Death is disappearance of an individual through competition and is dependent on the amount of neighbouring individuals: The two-dimensional domain is subdivided into fixed square interaction cells of area $\delta^2$. An individual disappears at rate $\lambda \cdot n$ when $n$ \emph{other} individuals are present in the same lattice cell. Thereby, $\lambda$ is a rate independent of $n$. The birth and death processes can be described by the binary interactions sketched in \FIG{fig:fig2}A and can be summarised as
\begin{align}
&X \xrightarrow{\mu} X + X
&X +X' \; (\text{inside } \delta^2) \xrightarrow{\lambda} X'\,.
\label{eq:birthdeath}
\end{align}
This choice of rules is also known as birth-coagulation process as disappearance occurs through coagulation \cite{DoeringPhysicaA03}.

In addition to birth and death, individuals are subject to dispersal in the form of a random walk, i.e., they diffuse in a two-dimensional continuous habitat with diffusion coefficient $D$, as depicted in \FIG{fig:fig2}A. This diffusive motion, together with the birth-death process, allows one to interpret the individual-based scheme as a discretised reaction-diffusion scheme (\SM{Appendix S1}).

All individual-based simulations are performed with a domain size of $1000 \times 1000$, an interaction range of $\delta = 1$, a diffusion coefficient of $D = 1$ and birth- and death rates of $\mu = 1$ and $\lambda=1$, respectively, unless specified otherwise.

When a band of individuals is set as initial condition, the system  invades the empty space with a fluctuating front propagating at an average constant speed controlled by the microscopic parameters and the associated level of demographic noise, which is larger for smaller density \cite{DoeringPhysicaA03, HallatschekPRL09}.

\section*{Single circular obstacles and hotspots}
\label{sec:oneobshot}

Inhomogeneities within which the microscopic parameters differ from their values outside are expected to shape the dynamics of the front. We refer to a patch that slows down or blocks the front as an `obstacle' and to a region that can be invaded faster than the surroundings as a `hotspot'.  

First, we study the effect of one single circular impermeable obstacle, realised by a locally vanishing diffusion coefficient, $D_2=0$. \FIG{fig:fig2}C shows a time series of a single realisation of an individual-based simulation as well as the average front obtained from many realisations (see \SM{Video S1} for all frames and \SM{Appendix S1} on how the front is determined). We observe that right after the front has passed the obstacle, a part of the front lags behind, resulting in a kink that then heals. This behaviour is in qualitative agreement with the observations of Ref.~\cite{MoebiusPLOSCB15}, where a `constant speed model' was used to describe front shape when a population front encounters an obstacle. In this model the front results from a collection of points that have the same distance to the initial front when taking into account the impermeability of the obstacle as sketched in \FIG{fig:fig2}B. The green line gives one example of a shortest path or `path of least time' between a point at the front and any point at the initial condition. The total front is constructed by finding all points that have the same distance to the initial front, see Ref.~\cite{MoebiusPLOSCB15} for details. In \FIG{fig:fig2}C we show this least-time front for a complete propagation around an obstacle (orange line). We observe that this construction recovers the average shape of the front from the simulations, including the kink, very well. The individual front is slightly lagging behind however as observed before \cite{MoebiusPLOSCB15}. Far away from the obstacle the front is well described by the envelope of two radial waves (dashed purple lines in \FIG{fig:fig2}C), initiated from the two vertical extremes of the obstacle and travelling with constant speed as will be discussed below.

The reverse situation of a `hotspot' can be achieved by setting the diffusion coefficient of individuals larger inside the inhomogeneity than in its surrounding. \FIG{fig:fig2}D shows the results of simulations where the diffusion coefficient inside the circular patch, $D_2$, is 2.5 times larger than outside (see \SM{Video S2} for all frames and \SM{Appendix S1} for details). The population expands faster within the hotspot, and a bulge forms to the right of the hotspot. The front dynamics can be described using a least-time consideration that assumes two different propagation speeds, $v_2$ and $v_1<v_2$, inside and outside the hotspot, respectively, see \FIG{fig:fig2}B. The front consists of the set of points whose paths back to initial condition are traversed in the same amount of minimal time (compared to alternative paths), in analogy to `Fermat's principle of least time' from classical optics \cite{HechtBook}. Using Snell's law, which can be derived from `Fermat's principle of least time', the resulting front dynamics can be obtained analytically (\SM{Appendix S2}, \SM{Fig. S1}). With $v_2 \approx 1.8 v_1$ (estimated from simulations of homogeneous systems), the resulting solution approximately captures the shape observed in the individual simulations (\FIG{fig:fig2}D, orange line). A combination of the planar front and a radial wave (dashed purple line in \FIG{fig:fig2}D), emitted from the centre of the hotspot describes front shape well far away from the hotspot, as will be explored below.

Overall, we find that a least-time description of front dynamics allows us to describe the dynamics of a population front encountering a single obstacle, within which diffusivity vanishes, or a single hotspot, a region where diffusivity is increased. Completely analogous observations are made when a population wave encounters a region with vanishing or increased birth rate (instead of diffusivity), see \SM{Fig. S3}.

\section*{Applicability of the least-time principle}

Before applying the least-time approach to more complex shapes and heterogeneous environments, we briefly outline its range of validity.
The validity of the least-time description relies on the possibility of replacing the dynamics of the whole population by an interface propagating orthogonal to itself with a locally-defined speed, i.e., a speed that depends on location only and not on, for example, direction or front dynamics at earlier times.

Population fronts are characterised by a transition from an unstable to a stable state. At a specified location, this transition occurs over a finite time. Spatially, this transition presents itself as a finite steepness of the front which corresponds to a front width. Note that this is a different measure to front or interface roughness, which can also be referred to as width. The coarsened, least-time approach implicitly assumes vanishing width. For the least-time approach to be a good description of the full dynamics, widths of traveling fronts firstly need to be very small compared to length scales of the system, i.e., the typical size of obstacles and hotspots or spacings between them.

Secondly, transient regimes are expected when a population encounters a hotspot or leaves it behind as the front does not instantaneously changes speed and steepness as it passes from one type of environment to another. These associated times are required to be negligible with respect to the time the front takes to pass through the hotspot. This condition is difficult to quantify, but can always be met by sufficiently large scales.

Thirdly, local front curvature is expected to have an effect on front speed in the underlying microscopic model \cite{MurrayBook} not reflected in the coarsened model where front speed is a purely local parameter. This effect can be important at the corner of an obstacle \cite{MoebiusPLOSCB15}, at the entrance of a hotspot, or at the kinks of perturbed fronts. Although the least-time approach does not capture these subtleties, their relative effect is expected to be small for large features.

We stress that individual-based models are particularly suitable to the study of heterogeneous media, since the presence of a natural cut-off (due to the discreteness of the individuals) leads to a unique and stable front speed \cite{vanSaarloosPR03}.

Finally, individual-based models are characterised by a natural roughness due to the stochastic nature of the growth process \cite{Halpin-HealyPR95}. In this paper we consider situations in which the size of the feature and the perturbation to the front by an obstacle or a hotspot are large compared to the typical scale of the roughness. For hotspots, this criterion depends not only on its size but also on its strength.

\section*{Obstacle width and hotspot length shape front at large distances}

The least-time considerations can be used to uncover which aspects of an obstacle's or hotspot's shape dictate front shape far away from the feature. \FIG{fig:fig2}B shows that the front in the shadow of the obstacle is associated with paths originating from an area around the obstacle's maximum width. These paths are the shortest path back to the initial front (compare also Ref.~\cite{MoebiusPLOSCB15}). \SM{Fig. S2A} depicts the front further downstream, highlighting this observation. This suggests that (i) the exact shape of the obstacle does not matter for the front shape far downstream and that (ii) two radial fronts, each originating from the widest part of the obstacle, describe the solution for general obstacle shapes at large distances downstream.

To test these arguments, we determined the fronts numerically for more general shapes, employing the fact that the least-time consideration is equivalent to the Eikonal equation,
\begin{equation}
\left| \nabla T(\vec{x})\right|=1/v(\vec{x}), \label{eq:Eikonal}
\end{equation}
which connects the arrival time $T(\vec{x})$ to the local front speed $v(\vec{x})$. Front shapes at different times are given by contours in the arrival time $T(\vec{x})$, which can be numerically obtained using the Fast Marching Method \cite{SethianPNAS96,scikit-fmmURL}.

\begin{figure}
\includegraphics[width=\textwidth]{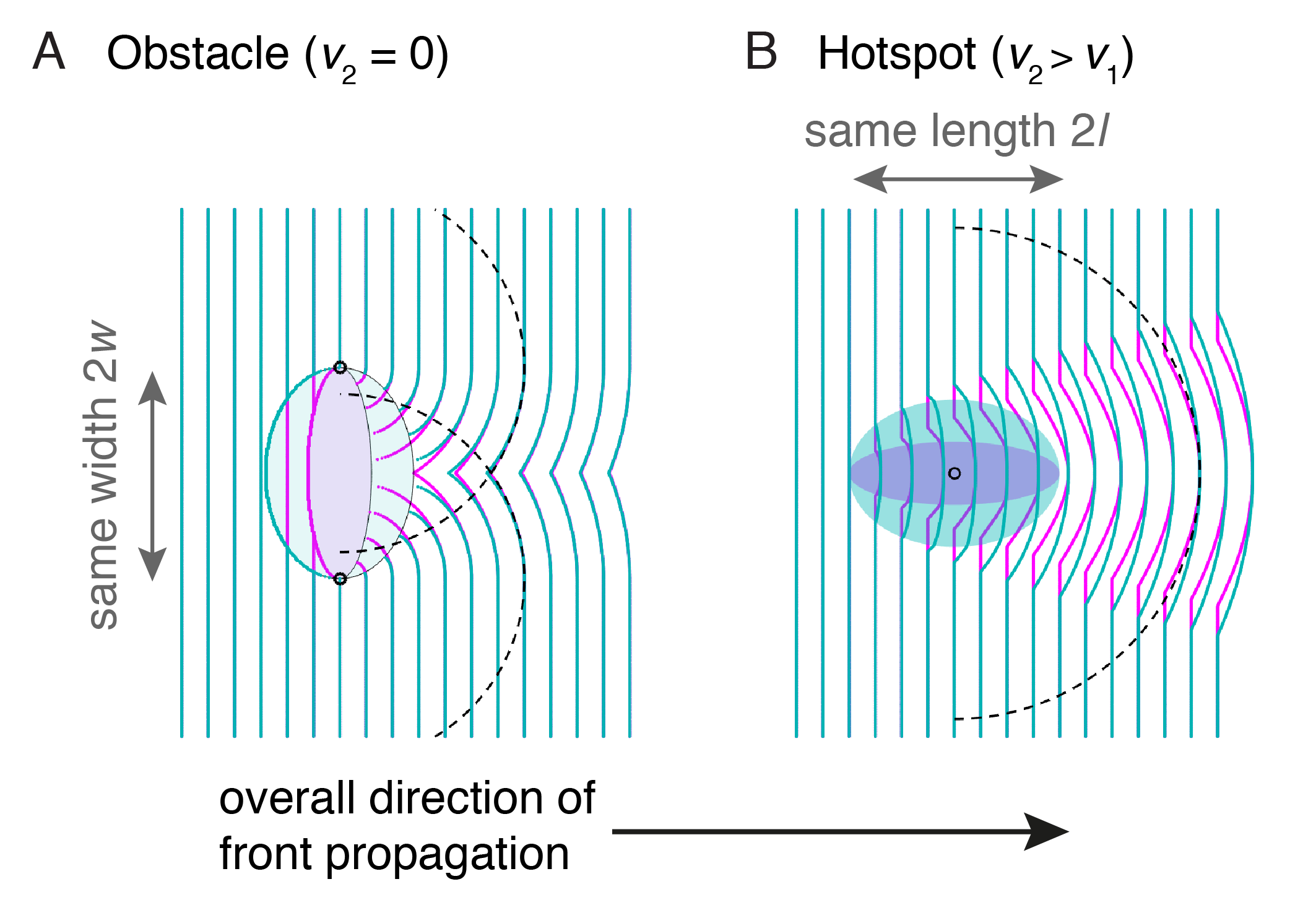}
\caption{Front shape before and after encountering a obstacle and hotspot, respectively, with front speed $v_1$ outside the feature and $v_2$ inside. \textbf{(A)} Solutions of Eikonal equation (magenta and cyan lines) at different times relative to elliptical obstacles ($v_2=0$, magenta and cyan-shadowed ellipses) with equal widths but varying aspect ratio. Half-circles originating from the sides of the obstacle (dashed lines with origins marked by black circles) capture front shape downstream from the obstacle. \textbf{(B)} Numerical solutions relative to elliptical hotspots ($v_2/v_1=1.2$, magenta and cyan ellipses) with equal length but varying aspect ratio. A half circle originating at the centres of the hotspots with radius given by Eq.~\eqref{eq:farfieldhot} describes the bulge in the front downstream from the hotspots.}
\label{fig:fig3}
\end{figure}

We chose two different elliptical obstacles with the same width, but different lengths, and computed the front numerically as depicted in \FIG{fig:fig3}A. Indeed, we observe that obstacles with the same width perturb the front far away from the obstacle in the same manner. The far-distance solution, constituted by two half circles, matches the numerical solution very well. To illustrate that the effect is not limited to convex shapes, we repeated the computation for a tulip-shaped obstacle, see \SM{Fig. S4}A, and again observe very good agreement.

The healing of the kink induced by the obstacle can be quantified by the opening angle $\theta$ and the indent size $s$ indicated in \FIG{fig:fig2}B. For large distances traveled since the obstacle was encountered we obtain:
\begin{equation}
\theta \approx \pi - \frac{2w}{d},\quad s \approx \frac{w^2}{2d}, \label{eq:farfieldobs}
\end{equation}
where $w$ is the half-width of the obstacle (equal to radius for circular obstacle) and $d$ is the distance traveled since the front has passed the point of maximum width. The size $s$ of the perturbation decays with the distance $d$ from the obstacle. See Appendix S1 of Ref.~\cite{MoebiusPLOSCB15} for a derivation.

Similar reasoning applies to the case of hotspots. \FIG{fig:fig2}B and \SM{Fig. S2B}  display the front behind a circular hotspot, together with the paths back to the initial front. Most paths from the bulge to the initial front pass through the central region of the hotspot, implying that the hotspot length is important for front shape far behind the hotspot. Numerically, we find that two ellipses with equal length, but different width, result in very similar bulges of the front as shown in \FIG{fig:fig3}B.

We find the bulge to be heuristically well described by a radial wave originating at the hotspot's centre and whose radius is given by
\begin{equation}
r = d+k, \quad\textrm{with}\quad k=2\,l\left (1-\frac{v_1}{v_2}\right ), \label{eq:farfieldhot}
\end{equation}
where $d$ is the distance between the unperturbed front and the centre of the hotspot and $l$ is the half-length of the hotspot (equal to radius for a circular hotspot). $v_1$ and $v_2$ are the front speeds surrounding and within the hotspot, respectively. This heuristic solution describes the bulge originating from a circular hotspot (\FIG{fig:fig2}D), elliptical hotspots (\FIG{fig:fig3}B), and even a tulip-shaped hotspot (\SM{Fig. S4}B) reasonably well. At its tip, the bulge proceeds the otherwise planar front by $k$ defined above corresponding to the advance a virtual marker gains by passing through the hotspot along the axis of symmetry. Note that $k$ does not depend on $d$, the distance traveled since the hotspot was encountered.

In the following we will refer to the approximate solutions far away from the obstacle (the two emitted radial waves from the extreme borders) and hotspot (the emitted radial wave from the centre of the hotspot) as `far-distance solutions' keeping in mind the heuristic nature for the case of hotspots.

Using the Eikonal equation \eqref{eq:Eikonal}, and the equations characterising the far-distance solutions \eqref{eq:farfieldobs} and \eqref{eq:farfieldhot}, one can illustrate an additional important property of the least-time description. If the environment including the obstacle or hotspot is stretched in all directions by the same factor, while front speed is kept constant, the arrival time is increased by the same factor, giving rise to a similarity solution of the front shape. This highlights that the findings presented can be applied at very different spatial scales, independent from the underlying mechanism of front propagation as long as the least-time principle can be applied. If this is not the case, for example if obstacles and hotspots are smaller than the characteristic front width discussed above, we expect a different front dynamics.

Taken together, we have seen how a least-time description of front propagation can predict the front computed with an individual-based simulation. This perspective allows us to characterise the perturbations induced by obstacles and hotspots, in particular the description as a superposition of the initial front with one or two radial waves, anticipated in \FIG{fig:fig2}C\&D (dashed purple lines). In the following, we will use these findings to investigate the effect of \textit{multiple} obstacles and hotspots on systems too large to be investigated by individual-based simulations.

\section*{Multiple obstacles and hotspots - a scattering process}

How are the perturbations by single obstacles and hotspots affected by other features downstream? Or, conversely, how is the effect of a feature influenced by perturbations upstream? To answer these questions, we will first consider a dilute regime employing the findings for individual obstacles and hotspots before investigating the regime of a dense pattern of features.

\begin{figure}
\includegraphics[width=0.9\linewidth]{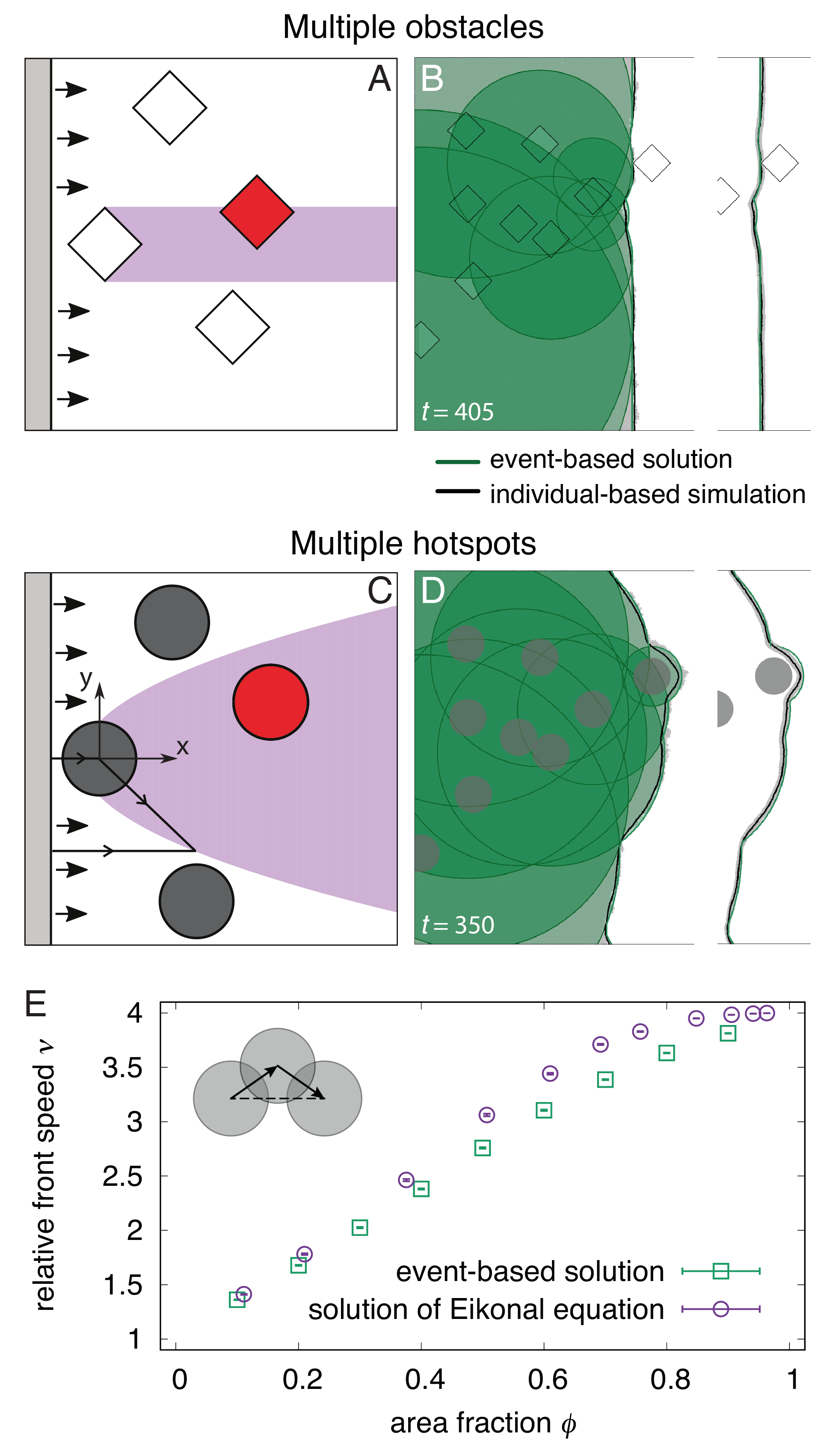}
\caption{\textbf{(A)} Region within which an obstacle perturbs the front. The purple area illustrates the shadow of the first of four obstacles: Obstacles inside this area (such as the red rhombus) will be affected by the perturbations created by the first obstacle. \textbf{(B)} Comparison of the event-based solution (green circles) with the result of individual-based simulations (grey dots for one realisation) in the presence of rhombus-shaped obstacles (half-width $w=50$). The envelope of the circles, determined by the event-based solution, matches well with the average front derived from many individual-based simulations (black line).
\textbf{(C)} and \textbf{(D)} Similar to panel (A) and (B), but now for circular hotspots with radius $R = 50$ and $D_2=10$ with radial waves originating from hotspots' centres. \textbf{(E)} Speedup of front $\nu$ (ratio of front speed in presence of hotspots and front speed outside of hotspots) computed with the event-based approach (green squares) and the numerical solution to the Eikonal equation (violet circles) for variable area fractions of hotspots $\phi$. Inset: Sketch of overlapping hotspots. The path through hotspots centres is longer than the shortest path between the leftmost and rightmost hotspot (dashed line).}
\label{fig:fig4}
\end{figure}

\FIG{fig:fig4}A displays four obstacles encountered by an original planar front. The purple region indicates the `shadow' of the obstacle, i.e., the area influenced by the first obstacle encountered. Only the obstacle overlapping with this region, shown in red, interacts with the perturbation created upstream, causing a more complex perturbation, because the red obstacle is reached by a non-planar front. For rhombus-shaped obstacles considered here, the front in their shadow is completely described by the radial waves discussed above, i.e., each corner of a rhombus acts as a `scattering point' from which a radial wave originates.

Front propagation in an environment with rhombus-shaped obstacles reduces to repeated scattering at the corners of rhombuses resulting in an `event-based solution'. The front is then constituted by the maximum (or envelope) of all radial waves (and the unperturbed planar front) which are not blocked by obstacles. \FIG{fig:fig4}B illustrates the success of this approach: The black line indicates the average front derived from the microscopic individual-based model which agrees with the envelope of the green circles, the event-based solution, after a few rhombuses have been encountered by the front. While for rhombuses, this scattering algorithm is exact, smooth curved boundaries would be associated with an infinite number of scattering events making this approach computationally unfeasible.

The perturbations induced by hotspots accumulate differently. The effect of a hotspot is not only felt in its geometrical shadow, but in a widening region as is evident from \FIG{fig:fig3}B. Using the heuristic approximation described above, the interaction region can be obtained by equating the distance $d$ a planar front would travel after passing the hotspot's centre with a radial wave of radius $d+k$ with $k$ defined in Eq.~\eqref{eq:farfieldhot}. The result is a sideways parabola, in the $x-y$ reference system with origin at the centre of the hotspot,
 \begin{equation}
 y=\pm \sqrt{k^2+2kx}\, .
 \label{eq:parabola}
\end{equation}
In \FIG{fig:fig4}C, the red hotspot, located within the parabola, will further accelerate the front, while hotspots indicated in dark grey are expected to advance the front independently.

The effect of several hotspots can be pictured as a succession of activation events: Each hotspot encountered by the front is `activated' and a radial wave originates from its centre. The planar wave and all radial waves can activate hotspots. The specific rules reflect that the front propagates with speed $v_2$ inside and speed $v_1$ outside the hotspot, respectively. The front is given by the envelope of all these individual circular waves and the initial planar front (see \SM{Appendix S1} for a detailed description). \FIG{fig:fig4}D illustrates this approach (green circles) and shows good agreement with the front determined from the individual-based simulation (black line). Since the event-based algorithm for hotspots uses the heuristic solution for large distances, we expect the resulting front to generally deviate from the exact solution.

Front speed is a key observable for the spatial spread of populations and the observable focused on below. We therefore use front speed to quantify the deviation between the event-based solution and the exact solution obtained by solving the Eikonal equation numerically, introduced above and described in \SM{Appendix S1}. Front speed is derived from the (mean) front position, defined as front position averaged along the direction vertical to the advancing front (averaged in vertical direction in \FIG{fig:fig4}) and reported relative to front speed in the absence of obstacles or hotspots, i.e., relative to $v_1$.

\FIG{fig:fig4}E displays this relative front speed $\nu$ in the presence of random hotspot configurations at variable area fractions $\phi$, i.e., different fractions of area covered by hotspots. Front speed derived from the event-based solution appears in good agreement with that from solving the Eikonal equation for small hotspot area fractions of up to $\phi \approx 0.3$. For intermediate area fractions of $\phi \approx 0.6$, front speeds obtained with both approaches deviate from each other significantly. At very high area fractions, both approaches result in an effective speed close to the speed expected in an environment fully covered with hotspots ($\nu=4$ for $v_2/v_1=4$). In general, the event-based approach underestimates front speed because in the event-based solution only paths through hotspot centres are considered even though shorter paths may exist (inset to \FIG{fig:fig4}E). This effect plays a minor role in the dilute regime and in the regime of very dense hotspots: In the former case, we expect the heuristic solution to describe the front well. In the latter case many, potentially aligned, hotspots exist. Taken together, the event-based solution provides qualitative insight into front dynamics, it is not suited to compute front speed for general inhomogeneous environments.

Having established that the least-time principle can be used to describe front dynamics around isolated and small groups of inhomogeneities, we will use the numerical solutions to the Eikonal equation to explore the front dynamics for much larger systems, different shapes of obstacles and hotspots, and for a wide range of area fractions. While it is not necessary to run the individual-based simulations to expanding populations in those environments, it would also be prohibitively costly computationally.

\section*{Front speed as function of obstacle density and shape}

The picture of individual obstacles inducing scattering events leads to a number of predictions: Several obstacles located in each others' shadows perturb the front repeatedly and, if occurring at all parts of the front simultaneously, lead to an overall slow-down of the front. Since perturbations originating from single obstacles heal with increasing distance from the obstacle, the cumulative effect of perturbations becomes stronger if obstacles are closer, i.e., in a denser configuration.

To test these predictions, we numerically solved the Eikonal equation using the Fast Marching Method for elliptical obstacles in a system as large as computationally feasible. We investigated elliptical obstacles such as in \FIG{fig:fig3}A because they are arguably less idiosyncratic then rhombuses. Without loss of generality, we chose the spatial scale of obstacles to be of order $1$. With the lattice constant chosen to be $1/15$, each obstacle is represented by hundreds of lattice sites. The width of the channel with periodic boundary conditions is set to $50$ and the length to $1300$, see \SM{Appendix S1} for more details. We computed the front dynamics for a random placement of obstacles at a given number density $\rho$ and obtained the front speed by linear fits of front position vs. time as described in \SM{Appendix S1}.

\FIG{fig:fig5}A displays relative front speed $\nu$ as a function of area fraction $\phi$ for four different ellipses which differ in length and width as well as snapshots of obstacle configurations and resulting front shape. Area fraction is a function of the product of number density $\rho$ and the semimajor and semiminor axes $R_a$ and $R_b$, i.e., $\phi = 1-\exp(-\rho \pi R_a R_b)$ as easily derived by change of variables and the well-known result for overlapping disks \cite{TorquatoBook}. As expected, for circular obstacles (purple and cyan symbols), front speed does not depend on the radius because the obstacle and front shape can be scaled with the same factor, as discussed above. At equal area fraction and for ellipses with same aspect ratio (green and orange symbols) front speed is more reduced if the long axis is parallel to the front, in agreement with our finding that it is the obstacle's cross section that is responsible for the front perturbation. However, this observation also implies that for a given environment, with aligned obstacles, front speed can depend on the angle of incidence and the environment can therefore be anisotropic with respect to front propagation.

We had seen that the `far-distance solution' of the front behind an obstacle is characterised by the obstacle’s width $b$ (but not its shape). Because the speed does not change when stretching the environment in all dimensions equally, speed cannot depend on $b$ alone or on non-dimensionless combinations of $b$ and the number density $\rho$, but instead should be a function of the dimensionless parameter $\rho b^2$. We indeed observe a strong dependency of speed $\nu$ on $\rho b^2$ (\FIG{fig:fig5}B). However, this dependency is imperfect: The front is slower for larger length to width ratios (i.e., for `longer' ellipses). This is because for `longer' ellipses, a larger fraction of the area is covered by obstacles increasing the path length. Conversely, a (hypothetical) system of thin rod-shaped obstacles is expected to provide an \textit{upper limit for front speed} in an environment of obstacles with width $b$ and number density $\rho$.

\begin{figure*}
\includegraphics{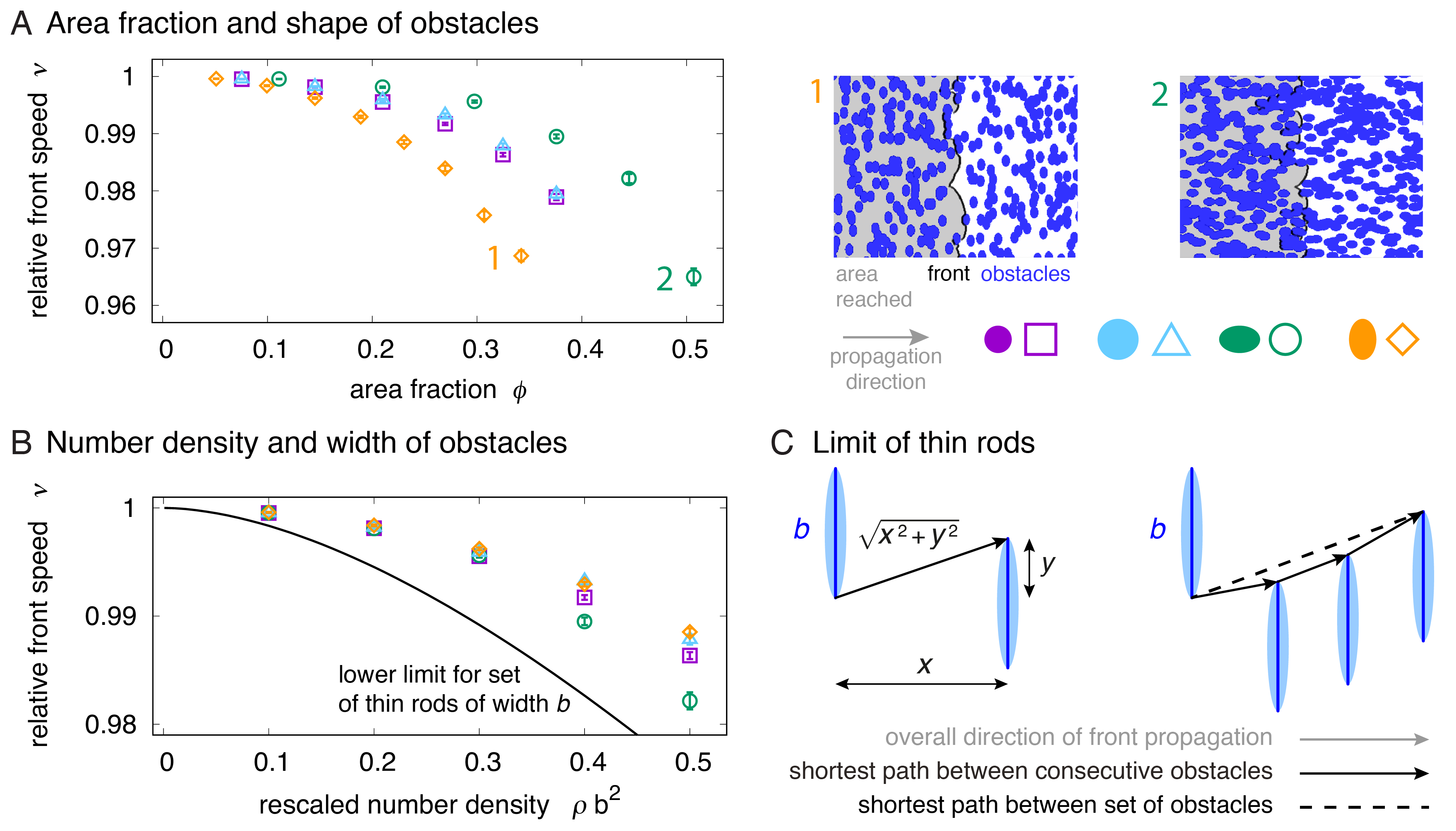}
\caption{Effect of many randomly placed obstacles on front speed. \textbf{(A)} (Left) Slow-down of front quantified by relative front speed $\nu$ obtained by numerically solving the Eikonal equation as a function of $\phi$, the area fraction covered by obstacles. Symbols and colours indicate ellipse-shaped obstacles with varying aspect ratio aspect ratio (purple \& blue: $1$, green: $3/2$, orange: $2/3$). (Right) Snapshots of obstacle configurations and resulting front shape from the numerical solution for parameters indicated on left. See \SM{Video S3} and \SM{Video S4} for corresponding videos of front movement. \textbf{(B)} Relative front speed $\nu$ for elliptical obstacles defined and computed as in (A), but as function of $\rho b^2$, with $\rho$ the number density of obstacles and $b$ the obstacle width. In addition, a lower limit for front speed in the presence of a system of thin rods is shown (black line, see \SM{Appendix S2}). \textbf{(C)} (Left) A sketch of the shortest path between the two rods or very elongated ellipses of width $b$, distance $x$, and overlap $y$. The relative increase in path length is given by $\sqrt{x^2 + y^2}/x$. (Right) A sketch of possible paths through a geometry with randomly distributed parallel rods. The dashed line shows the absolute shortest path, the solid arrows show the path constructed from the shortest path between consecutive obstacles.}
\label{fig:fig5}
\end{figure*}

A system of very thin rods lends itself to an understanding of the cause for the slow down. When the projections of rods in the direction of front propagation overlap, the propagation path will graze the corners of the rods (similar to the `scattering description' we employed above). The slow-down of the front is then given by the increase in path length, relative to the straight path in the propagation direction, as shown in \FIG{fig:fig5}C. The path grazing the corners of all overlapping rods is however not always the shortest path as evident in the exemplary configuration in \FIG{fig:fig5}C. There, the shortest path directly connects the first and the last rod (dashed line), while considering nearest neighbours (solid arrows) constructs a longer path that connects all rods in between. Assuming the path grazes all consecutive overlapping rods, allows one to compute front speed analytically by integrating over all possible overlapping rod pairs (see \SM{Appendix S2}). The result (black line in \FIG{fig:fig5}A) is a \textit{lower limit for front speed} for a system of thin rods.

Taken together, we derived a lower limit for front speed in a system of rods of length $b$ which itself sets an upper limit for sets of obstacles of width $b$. This finding is nevertheless useful as it sheds light on why obstacles reduce front speed only marginally. It is not the area covered by obstacles that sets front speed, but the extension of path length that is required to graze the corners of (a subset of) obstacles.

However, with increasing density, the shape of the obstacles becomes important and obstacles may overlap more often. Since the front cannot propagate inside obstacles, the front will stop when so many obstacles overlap in transversal direction that no unobstructed path exists. Such blockages can arise in finite domains even at a filling fraction smaller than the critical percolation threshold, which is for circular obstacles in a infinite system given by $\phi\approx 0.68$ \cite{QuintanillaJPA00,TorquatoBook}. We have limited our analysis to significantly lower area fractions, for which statistics on the front speed can still be easily acquired. We expect the front to slow down dramatically close to or above the percolation threshold. This slow-down has been addressed recently in lattice-based growth models \cite{MogliaPRE16,GralkaELife19}.

\section*{Front speed as function of hotspot density, shape, and intensity}

\begin{figure*}
\includegraphics{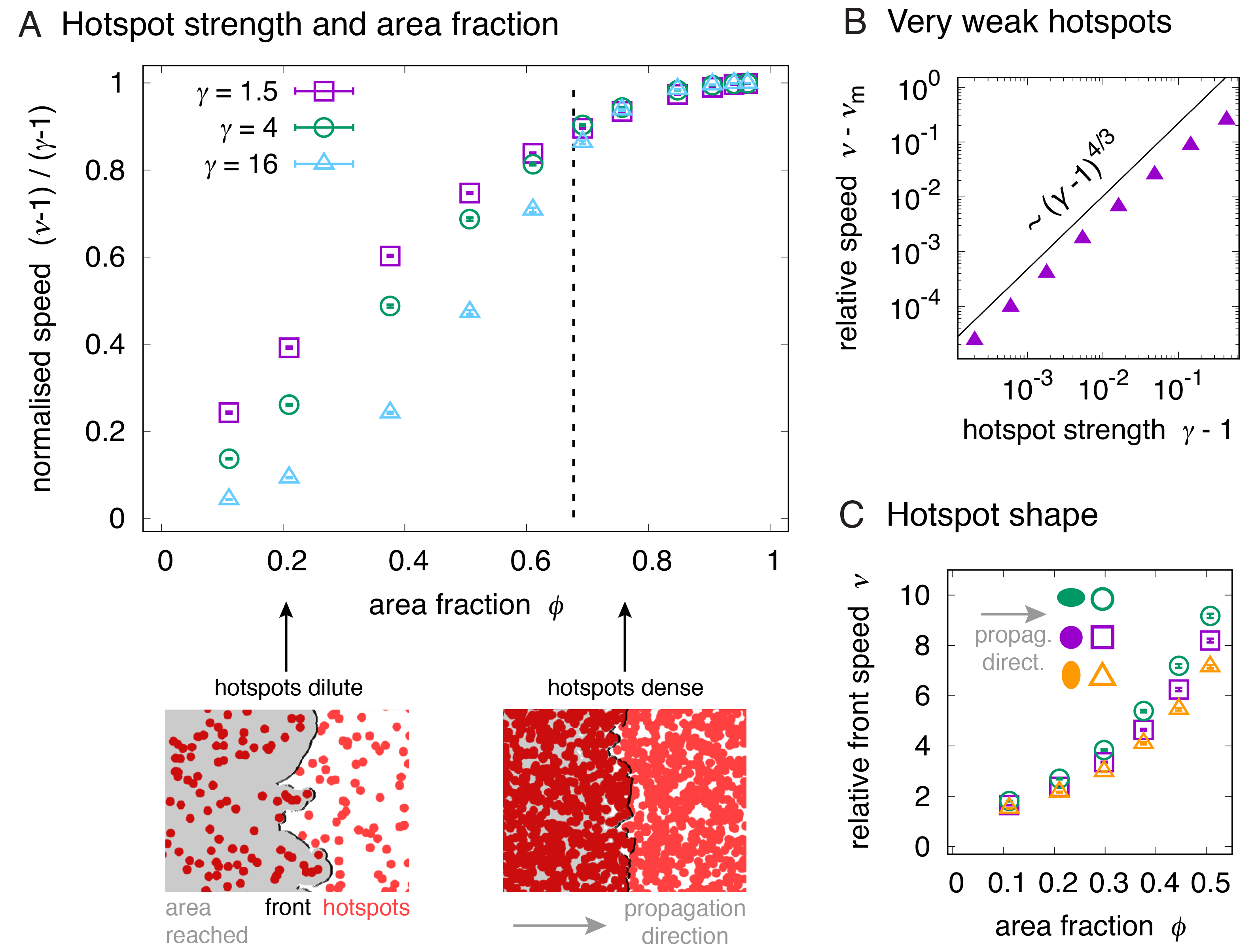}
\caption{Effect of many randomly placed hotspots on front speed, obtained by numerically solving the Eikonal equation. \textbf{(A)} (Top) Normalised front speed $(\nu-1)/(\gamma-1)$ as function of hotspot strength $\gamma$ and area fraction $\phi$. The dashed vertical line corresponds to the percolation threshold in an infinite system ($\phi=0.68$). See \SM{Fig. S5} for the same numerical results, plotted relative to the weighted harmonic mean of local front speeds. (Bottom) Snapshots of dilute and dense hotspot configurations and resulting front shape. See \SM{Video S5} and \SM{Video S6} for corresponding videos of front movement. \textbf{(B)} Relative speed $\nu-\nu_m$ with $\nu_m=\phi\gamma+(1-\phi)$, the spatial average of local front speed. The solid line indicates a power law $\sim (\gamma-1)^{4/3}$. \textbf{(C)} Relative front speed $\nu$ as a function of area fraction $\phi$ for strong elliptical hotspots with strength $\gamma=16$ for three different aspect ratios (purple: $1$, green: $3/2$, orange: $2/3$). See \SM{Fig. S6} for an equivalent plot, but for weak hotspots with $\gamma=1.5$.}
\label{fig:fig6}
\end{figure*}

A single hotspot leads to a transient increase in local front speed, resulting in a bulge with constant size in the direction of front movement and sideways spreading along the front (\FIG{fig:fig3}B). We therefore expect multiple hotspots to result in an overall speed-up of the population front. We first consider the case of circular hotspots with intensity $\gamma=v_2/v_1$ and area fraction $\phi$. \FIG{fig:fig6}A depicts the speed-up as obtained from solving the Eikonal equation. Relative front speed $\nu$ is plotted as $(\nu-1)/(\gamma-1)$, which varies between $0$ and $1$ for any $\gamma$ and any $\phi$ between $0$ and $1$.

The shape of the speed-up $\nu(\phi)$ depends on hotspot strength $\gamma$. While for small $\gamma$, it resembles a concave function, we observe a sigmoidal shape for large $\gamma$ with the point of inflection at an intermediate area fraction below the percolation threshold ($\phi\approx 0.68$ for an infinite system). We hypothesise that the larger slope at intermediate $\phi$ is due to a change in how the front is sped up when increasing $\phi$: While for a dilute system, the front is locally accelerated by individual hotspots (\FIG{fig:fig6}A, \SM{Video S5}), for large area fractions the hotspots constitute a connected path and the effective front speed depends on the length of this percolating path (\FIG{fig:fig6}A, \SM{Video S6}). In a finite domain, percolation can occur below or above the percolation threshold in the thermodynamic limit, depending on the actual hotspot configuration. We expect this fact to be reflected in a larger variance in the measured speed, $\langle (\nu - \langle \nu \rangle)^2 \rangle$, close to the critical area fraction.  

In a simple linear habitat, as sketched in \FIG{fig:fig1}B, the front speed along this linear path does not depend on the arrangement of hotspots, but solely on the area fraction $\phi$. The relative front speed, $\nu$, is given by the weighted harmonic mean, $\nu_h=\left(\phi/\gamma+(1-\phi)\right)^{-1}$ (\SM{Appendix S2}). This result is a lower bound for the front speed-up in two dimensional systems, since in the latter many more paths with possibly shorter travel times exist, in addition to a straight path mimicking a linear habitat. Indeed, $(\nu-1)/(\nu_h-1)$, depicted in \SM{Fig. S5}, is larger than one for all area fractions. It is largest around the percolation threshold and for large hotspot strength.

For $\gamma \approx 1$, i.e., very weak hotspots, the results from scaling \cite{KersteinPRL92}, numerical \cite{KersteinPRE94}, and mathematical analysis \cite{MayoPLA07} of the speed-up of a Huygens front in isotropic random media apply to our system. In particular, we expect the speed-up minus the relative spatial average  of local front speed, $\nu_m=\phi \gamma + (1-\phi)$, to scale with the strength of the perturbation, $\gamma-1$, as 
\begin{equation}
\nu - \nu_m \propto (\gamma-1) ^{4/3}.
\label{eq:MayoKersteinpred}
\end{equation}
\FIG{fig:fig6}B is consistent with this prediction for $\phi=0.5$.

So far, we have considered circular hotspots and addressed the dependence of the speed-up on their intensity and area fraction. As discussed above, the length of an individual hotspot determines much of the front shape downstream. In particular, ellipses with equal length but different aspect ratio result in very similar front shapes (\FIG{fig:fig3}B). Conversely, we expect that ensembles of longer hotspots speed up the front more than ensembles of wider hotspots at equal area fraction. Numerical solution confirms these predictions, see \FIG{fig:fig6}C for strong hotspots and \SM{Fig. S6} for weak hotspots of varying aspect ratio.

\section*{Discussion}
The effect of inhomogeneities on population fronts depends on the type of inhomogeneities perturbing the front. Both classes of features considered here, obstacles and hotspots, perturb the population front in their own distinct way: The kink caused by an obstacle is transient and limited to the obstacle's width. Hotspots create a permanent perturbation that spreads along the front. Both effects can readily be understood by least-time arguments and analogies to geometrical optics at sufficiently large scales. Far from the inhomogeneity, the front can be described as a combination of radial waves induced from the outer corners of an obstacle or from the centre of a hotspot, respectively, which paints a picture of front propagation by repeated scattering events in environments with many inhomogeneities. On the quantitative side, the front speed can be obtained numerically using the Fast Marching Method, i.e., by solving the Eikonal equation. This allowed us to investigate dependence of front speed on the environment's parameters such as area fraction of the features' shape.

\begin{figure}
\includegraphics[width=\linewidth]{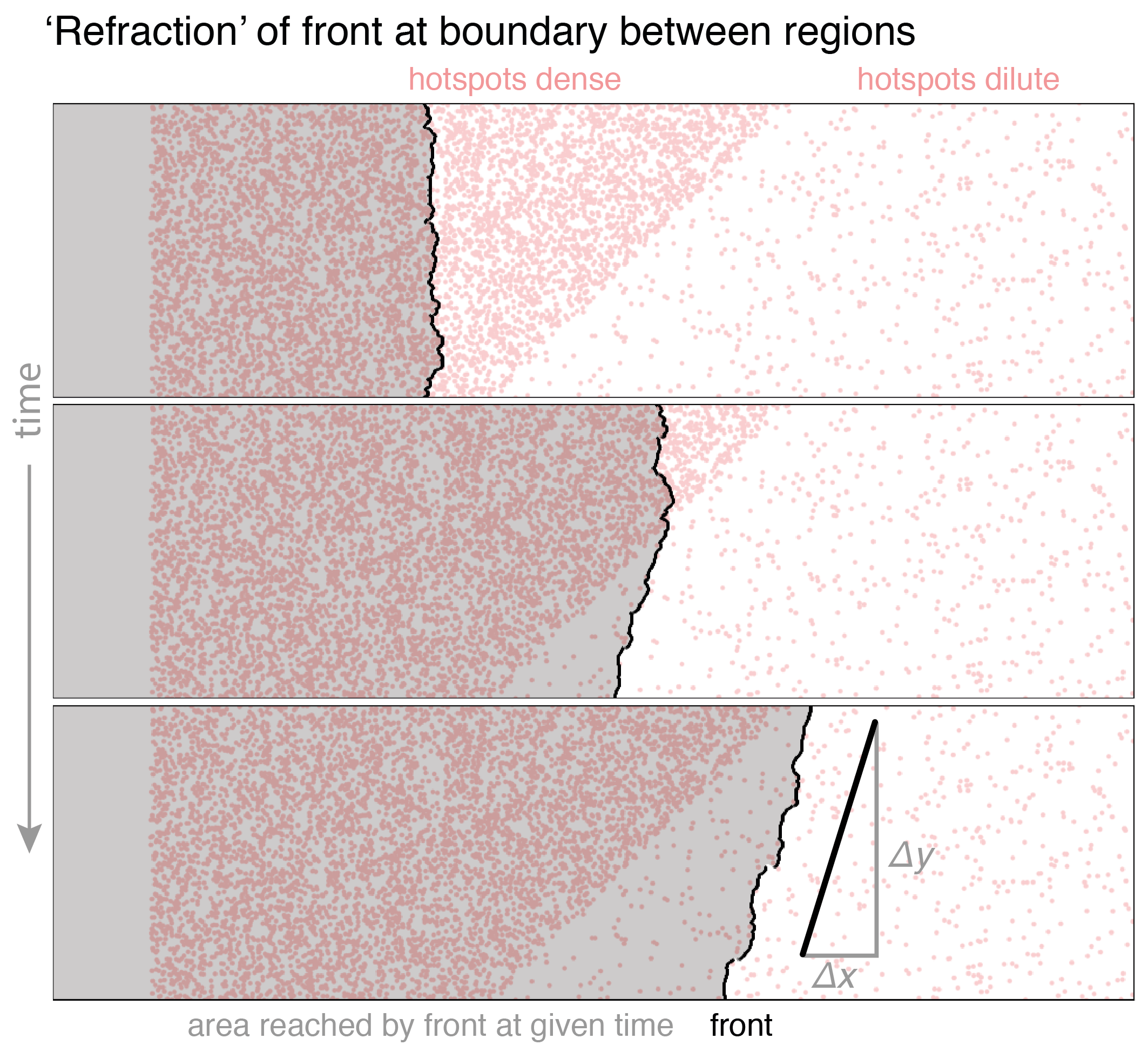}
\caption{`Refraction' of a front, obtained by numerically solving the Eikonal equatoin, at an interface between a region with dense ($\rho=0.150$) and dilute hotspots ($\rho=0.015$) of strength $\gamma=2.0$, tilted at 45\degree relative to the initial front. Upon encountering the interface, the front changes overall direction, manifesting in a tilt. The tilt angle is in agreement with the prediction based on the measured front speed at the area fractions of hotspots to the left and right using Snell's law \cite{HechtBook} ($v_\textrm{left}=1.74$, $v_\textrm{right}=1.14$, $\Delta y / \Delta x = \tan\left(\pi/4+\arcsin\left(v_\textrm{right}/(\sqrt{2}\,v_\textrm{left})\right)\right)$. We attribute deviations to boundary effects at the top and bottom of the channel. See \SM{Video S7} for a video of the full solution.}
\label{fig:fig7}
\end{figure}

The least-time description and the Eikonal equation employed here also arise in geometrical optics. Intuition gained from studying optics carries over to a large extent. To push the analogy further onto larger length scales, let us consider two areas with different hotspot density placed next to each other, with the interface tilted by 45\degree with respect to the initial front direction as illustrated in \FIG{fig:fig7}. From \FIG{fig:fig6}A we expect the front to propagate faster at high than at low hotspot density - and thus refraction of the front at the interface. Indeed, \FIG{fig:fig7} illustrates that as the front transitions from the region with dense hotspots to the region with dilute hotspots, it changes overall direction. The refraction angle predicted from Snell's law with propagation speeds measured in analogous homogeneous systems matches the observed tilt of the front.

The analogy to geometrical optics arises whenever fronts propagate in normal direction with a locally-determined speed that is independent from, e.g., front curvature, and thus found its application in other fields such as the prediction of forest fire fronts \cite{PreislerInBook13}. However, the analogy with optical phenomena is limited. For example, constructive and destructive interference will not occur in population expansions considered here. Reflection, which can be derived from Fermat's principle of least time \cite{HechtBook}, cannot be observed, because populations always expand into empty domains.

A large body of literature has investigated the effects of heterogeneities in one-dimensional, in particular periodic, habitats, see, e.g., \cite{LutscherTPB20,AzimzadeSciRep20}. Our study highlights that the results for linear habitats are generally not transferable to higher dimensions and thus not to many scenarios in nature. In the case of obstacles embedded in a two-dimensional environment, stagnation of front propagation can only occur when the area fraction is around or above the percolation threshold and there is no `free path' available to propagate further. In the case of hotspots, propagation is faster than in a corresponding linear habitat since many more paths are available. Thus, two-dimensionality suppresses the effect of obstacles and intensifies the effect of hotspots. We limited ourselves to random ensembles of potentially overlapping features of equal shape and orientation. Numerous questions arise that may be topic for future research: (i) A system of dense hotspots can be interpreted as a system of dilute imperfect obstacles, which are not circular. Can we predict front speed in this regime building on the statistics of the complement of overlapping disks \cite{TorquatoBook}? (ii) If obstacles are placed such that open channels exist within with the front can propagate undisturbed, front speed is not affected. Can we better predict front speed by identifying these channels? (iii) What is the front dynamics in complementary environments such as those generated from fractals \cite{HodgsonPLOSONE12}?

We envision our findings to support the study of macroscopic invasions in two different ways: Firstly, if researchers find evidence that a population expansion is governed by spread in normal direction, they can follow our approach of numerically solving the Eikonal equation to make predictions for front position at later times. This is especially useful should they wish to predict the front in large systems or for a large number of different habitats which is not feasible with individual-based simulations. Secondly, we believe the intuition gained from geometrical arguments can be used to understand even those environments which do not fulfil the requirement of a local front speed.

The least-time considerations and the Eikonal equation are fully deterministic and cannot capture fluctuations present in a single realisation of a population front of discrete individuals such as illustrated in \FIG{fig:fig2}. While we found the average over many realisations to be well described by the deterministic least-time consideration, it is possible that fluctuations drive individual expansions into different overall front dynamics, a question that warrants further investigation. Relatedly, deterministic dynamics of the population front does not imply deterministic evolution of the expanding populations. Even if the population expands it range mostly deterministically, a small population size at the front and the associated large genetic drift lead to gene surfing and gene segregation \cite{EdmondsPNAS04,HallatschekPNAS07}. The evolutionary dynamics is thereby influenced by the shape and dynamics of the front \cite{HallatschekPNAS07,AliPRE13}. Previous work has shown how the effects of obstacles and bumps on the evolutionary dynamics can be understood using the dynamics of front shape \cite{MoebiusPLOSCB15,BellerEPL18}. In particular, there is a relationship between lineages, the set of locations of subsequent birth events, and the shortest path used to construct the front in the analytical solution \cite{BellerEPL18}. This suggests that this work, in particular the characterisation of paths of least time, might help understand the collective effect of many large obstacles or hotspots on the genetic composition of the invading population, complementing recent work that characterised lineages in disordered environments without spatial correlation \cite{GralkaELife19}.

\section*{Acknowledgements}
We thank the main developer of the Python module scikit-fmm, Jason Furtney, for incorporating periodic boundary conditions on our request and Bert Wuyts for comments on an earlier version of the manuscript. Many of the computations in this paper were run on the FASRC Cannon cluster supported by the FAS Division of Science Research Computing Group at Harvard University.

\section*{Funding statement}
This work is supported by the Stichting voor Fundamenteel Onderzoek der Materie (FOM), the Netherlands. This article is based upon work from COST Action MP1305, supported by COST (European Cooperation in Science and Technology). Work by DRN was supported by the National Science Foundation, through grants DMR-1608501 and DMR-1435999.

\section*{Data accessibility}
The source code corresponding to the event-based solution and the solution of the Eikonal equation is available on Zenodo, https://doi.org/10.5281/zenodo.5513567, and GitHub, https://github.com/wmoebius/inhomogeneities\_one2many.

\section*{Author contributions}
WM, FTe, and KMJA participated in the design of the study and interpretation of results, performed calculations and simulations, and drafted the manuscript. RB, DRN, and FTo participated in designing the study, interpreting results, and writing the manuscript. All authors gave final approval for publication. WM \& FTe contributed equally.

\FloatBarrier

\bibliography{bib_obstacles_hotspots}

\end{document}


\title{Supplementary Material: \\
The collective effect of finite-sized inhomogeneities on the spatial spread of populations in two dimensions}%

\author{Wolfram M\"obius}
\thanks{These two authors contributed equally.}
\author{Francesca Tesser}
\thanks{These two authors contributed equally.}
\author{Kim M. J. Alards}
\author{Roberto Benzi}
\author{David R. Nelson}
\author{Federico Toschi}

\maketitle

\section{Appendix: Simulation and Numerical Methods}

\subsection{Individual-based Simulation}

The population model used is based on a birth process (duplication of individuals), a death process (disappearance of individuals), and dispersal of individuals through diffusion as explained in the main text. Starting with a small population at the simulation domain's boundary, we simulate propagation of the population front into the empty domain. The discrete nature of the model results in the presence of a natural cut-off in the resulting concentration field, fluctuations in overall number of individuals, and fluctuations of the front.

Without loss of generality, we choose the front to propagate in $x$-direction. We impose periodic boundary conditions along $y$-direction and infinitely unfavourable conditions outside the domain along $x$-direction, i.e., individuals disappear from the system if passing the domain boundaries. Due to this loss of particles from the domain, persistence of a population in a finite-sized domain is not guaranteed. The conditions for persistence of the population have been studied in continuous and discrete systems \cite{RyabovMMNP08,BertiPRE15}. In our simulations, the initial domain occupied by the population is large enough so that the population always expands into the empty domain and persists for the duration of the simulation.

The same discretisation of the domain into squares of size $\delta^2$ used to determine the disappearance of particles is used to determine the front (see main text and Fig. 2A therein). For each window of edge length $\delta$ in $y$-direction specified by $y_i$, the front is defined by the particle furthest along the $x$-direction, resulting in a set of points $x_i(y_i)$. For a given simulation time, we then obtain the mean and standard deviation, which is either reported or used to obtain front speed.

For the set of reaction rules used, a macroscopic continuum equation for the concentration of individuals can be derived, as described in Ref. \cite{PigolottiTPB13} and \cite{PerlekarJPCS11}. The level of noise in the model is determined by $1/n_e$, $n_e=\sqrt{\mathcal{N}}\sqrt{D/\mu}$, where $\mathcal{N}$ is the typical individual density, so that $n_e$ is the size of the actual interacting population in one generation time. It can be shown that the deterministic FKPP equation is recovered in the no-noise limit, $n_e\rightarrow \infty$, where the propagation speed equals to $v_{\text{FKPP}}=2\sqrt{D\mu}$. An expression for the speed of the front is known both in the weak $n_e\gg1$ and strong $n_e\ll1$ noise limit \cite{ConlonJStatPhys05}. However, we are not aware of an analytical expression for the regime of intermediate level of noise.

Unless otherwise noted, we chose the following parameters: birth or duplication rate $\mu=1$; death or disappearance rate $\lambda=1$ (to be multiplied by number of \emph{other} particles within region of size $\delta^2$); diffusion coefficient $D=1$. The edge length of square lattice cells is set to $\delta=1$. The size of the domain is $1000\times 1000$ with boundary conditions as described above. The population is initially placed on a sharp band of width $10$. Either an individual feature is located in the centre of the domain or, in the case of multiple feature, the centres are positioned randomly. To estimate local front speed (e.g., within hotspots), we determine front speed in a homogeneous system for the given set of parameters.

\subsection{Event-based Approach}

Far downstream from an individual hotspot or obstacle encountered by a planar population front, the front can be described as a combination of the original front and a set of radially expanding fronts as explained in the main text. For rhombus-shaped obstacles these radial population waves are emitted from the corners on the side, while for circular hotspots the centre of the wave coincides with the centre of the hotspot. We can therefore regard the accumulation of these wave-like perturbations as an on-going scattering process. This results in an event-based solution for the front shape illustrated in Fig. 4B,D of the main text. The details depend on whether obstacles or hotspots are considered as detailed below. For clarity, we here describe a continuous-time algorithm, the algorithm implemented uses discrete time steps.

We consider rhombus-shaped obstacles, such that radial waves are emitted only from the four corners of the rhombus. We start our analysis from a linear unperturbed front that propagates through the domain with speed $v_1$. As soon as this front encounters one of the corners of an obstacle, a radial wave is emitted from this corner. All following scattering points can be activated either by the planar front or by waves emitted from active scattering points. The requirement for such activation event is that the scattering point can be reached by the planar front or the radial wave, i.e., that no obstacles are blocking the path back to the scattering point or initial front. At a specific time, the front is given by the envelope of all emitted waves and the unperturbed planar front as long as they are not blocked by obstacles.

The hotspots we consider are circular regions with radius $R$ and within which the front propagates with speed $v_2$, which is larger than the propagation speed $v_1$ outside the hotspot. Scattering occurs at the centres of the hotspots. Upon activation, a radial wave originates from the hotspots centre and advances with speed $v_2$ inside and speed $v_1$ outside the hotspot. Scattering points are activated when they are encountered either by the planar front or by a radial wave emitted from an already active scattering point. We thereby need to take into account that the wave propagates with speed $v_2$ inside hotspots to be activated and thus distinguish three different scenarios: (i) The scattering point is activated by the planar front. Here we have to take into account that the planar front propagates faster inside the hotspot, i.e., the hotspot is activated when the planar front has travelled a distance of $Rv_1/v_2$ inside the hotspot. (ii) A scattering point is activated by the radial wave of an active scattering point whose centre is at least a distance $2R$ away. In this case, we take into account that the radial wave travels faster inside the two hotspots. (iii) Activation can occur by a radial wave originating from a hotspot overlapping with the hotspots of interest. In this case the distance between the scattering points is smaller than $2R$ and the complete path is travelled with speed $v_2$. At a specific time, the front is the envelope of the original planar front and the radial waves of all activated scattering points.

The corresponding source code is available on Zenodo, https://doi.org/10.5281/zenodo.5513567, and GitHub, https://github.com/wmoebius/inhomogeneities\_one2many.

\subsection{Solving the Eikonal equation using the Fast Marching Method}

To numerically determine a front whose time evolution is governed by the principle of least time we numerically solve the Eikonal equation $\left| \nabla T(\vec{x})\right|=1/v(\vec{x})$, which connects the (spatially varying) speed $v(\vec{x})$ to the arrival time $T(\vec{x})$. The front at time $t$ is given by contour lines of $T(\vec{x})$, i.e., the front consists of all $\vec{x}$ with $T(\vec{x})=t$. For numerical reasons we chose a slightly different definition of the front as described below.

The Eikonal equation was solved numerically using the Fast Marching Method \cite{SethianPNAS96}, implemented in the Python module \texttt{scikit-fmm} version 2021.2.2 \cite{scikit-fmmURL}. In the following, we describe the parameters used to determine front speeds. Without loss of generality, we chose the size of obstacles and hotspots to be on the order of 1. Any other size and appropriate scaling of the remaining parameters would lead to the same solution of the Eikonal equation. The lattice constant for the numerics was set to $1/15$, i.e., each obstacle or hotspot is represented by a few hundred lattice sites. The front propagates along a channel of length $1300$ and width $50$ with periodic boundary conditions in the latter direction. This choice reflects a trade-off between computational feasibility and accuracy, see Figs. \ref{fig:figS7} and \ref{fig:figS8} for selected data computed with a finer lattice or a wider channel. 64 individual environments were simulated to infer front speeds, see below for details.
Obstacles or hotspots were placed randomly with size and shape as specified in the main text and figures. We used the relationship $\phi = 1-\exp(-\rho \pi R_a R_b)$ relating number density $\rho$ and area fraction $\phi$ with $R_a$ and $R_b$ the semimajor and semiminor axes of the ellipse in the thermodynamics limit, which can be derived from the well-known result for overlapping disks \cite{TorquatoBook} using a change of variables. To avoid overlap of the initial front with either obstacles or hotspots, we extended the channel to one side by length $50$ and placed the initial front at the far side. This region was also used to `roughen' the front through a set of hotspots before entering a region with very weak hotspots investigated in Fig. 6C.

Different parameters were used to illustrate the effect of `refraction' at the scale of the environment in Fig. 7 of the main text. Channel length was set to $500$ and channel width to $150$ (no periodic boundary conditions).

For Fig. 3 of the main text, we set the semi-major axis of the ellipses to $1$. The tulip of Fig. \ref{fig:figS4} has a length of $2$ and a width of $1$. Due to the much smaller domain size, we were able to set the lattice constant to $1/100$. 

The corresponding source code is available on Zenodo, https://doi.org/10.5281/zenodo.5513567, and GitHub, https://github.com/wmoebius/inhomogeneities\_one2many.

\subsection{Determining front shape and front speed when following the event-based approach and when solving the Eikonal equation}

The front dynamics is fully described by the time $T$ at which position $(x,y)$ is reached. The points constituting the front between times $t$ and $t+\delta t$ are given by all $(x,y)$ for which $t\le T(x,y)\le t+\delta t$. While this is conceptually straightforward, it can be numerically challenging. This is in particular true for perfect obstacles which have a final arrival time at their boundary, but whose interior can never be reached by the front. We therefore defined the front as
\begin{equation}
h(y,t)|_T=\max_{T(x,y)\le t} x\, .
\end{equation}
Note that the two ways to infer the front may result in different front shapes and thus different mean front positions $\bar h(t)$ and front roughness $w(t)$,
\begin{equation}
\bar h(t) = 1/L \int h(y,t) \,\textrm{d}y,\quad \quad w(t) = 1/L \int \left(h(y,t)-\bar h(t)\right)^2 \,\textrm{d}y\, .
\label{eq:frontwidth}
\end{equation}
where $L$ is the width of the channel the front is propagating in. However, the front speed, the main observable in this work, is unaffected once front dynamics has reached a steady state.

Since the Fast Marching Method is lattice-based, the integrals in Eq.~\ref{eq:frontwidth} were replaced by the appropriate sums. In the case of the event-based solution, a continuous curve is in principle accessible, but the front was discretised to apply the same analysis procedure as for the solutions of the Eikonal equation.

For analysis, we obtained $\bar h(t)$ and $w(t)$ for different realisations of the environment with the same parameters, but different configurations of randomly placed obstacles and hotspots. To obtain front speed we fitted a line to each $\bar h(t)$ in the range $850\le \bar h(t)\le 1050$ with slope indicating front speed. From that ensemble of front speeds we computed and report the mean and standard error of the mean for $64$ trajectories.

At the beginning, when the originally flat fronts encounters the obstacles or hotspots, there is a transition period within which the instantaneous speed and front width transition to the steady-state values. When determining the fitting range stated above, we used a plot of width and front speed as a function of how far the front had progressed to identify a suitable range used to determine front speed. To facilitate this transition from a flat to a rough front, we inserted a number of stronger hotspots in front of the channel with very weak hotspots investigated in Fig. 6C.

\section{Appendix: Analytical Results}

\subsection{Front shape for circular hotspots}

To compute front shape of a planar front encountering a circular hotspot we need to find the set of points which are reached at a given time $t_\textrm{total}$. To obtain this set of points analytically, we use that fact that shortest paths are composed of linear stretches outside and inside the hotspot, respectively. Deflection (refraction) occurs at the hotspot-background interface, which can be described by Snell's law. It relates the angle of incidence, $\theta_1$, and the angle of refraction, $\theta_2$, through the propagation speeds $v_1$ and $v_2$ as
\begin{equation}
\frac{\sin(\theta_1)}{\sin(\theta_2)} = \frac{v_1}{v_2}\, .\label{eq:Snell}
\end{equation}
\begin{figure}[t]
\includegraphics[width=0.65\textwidth]{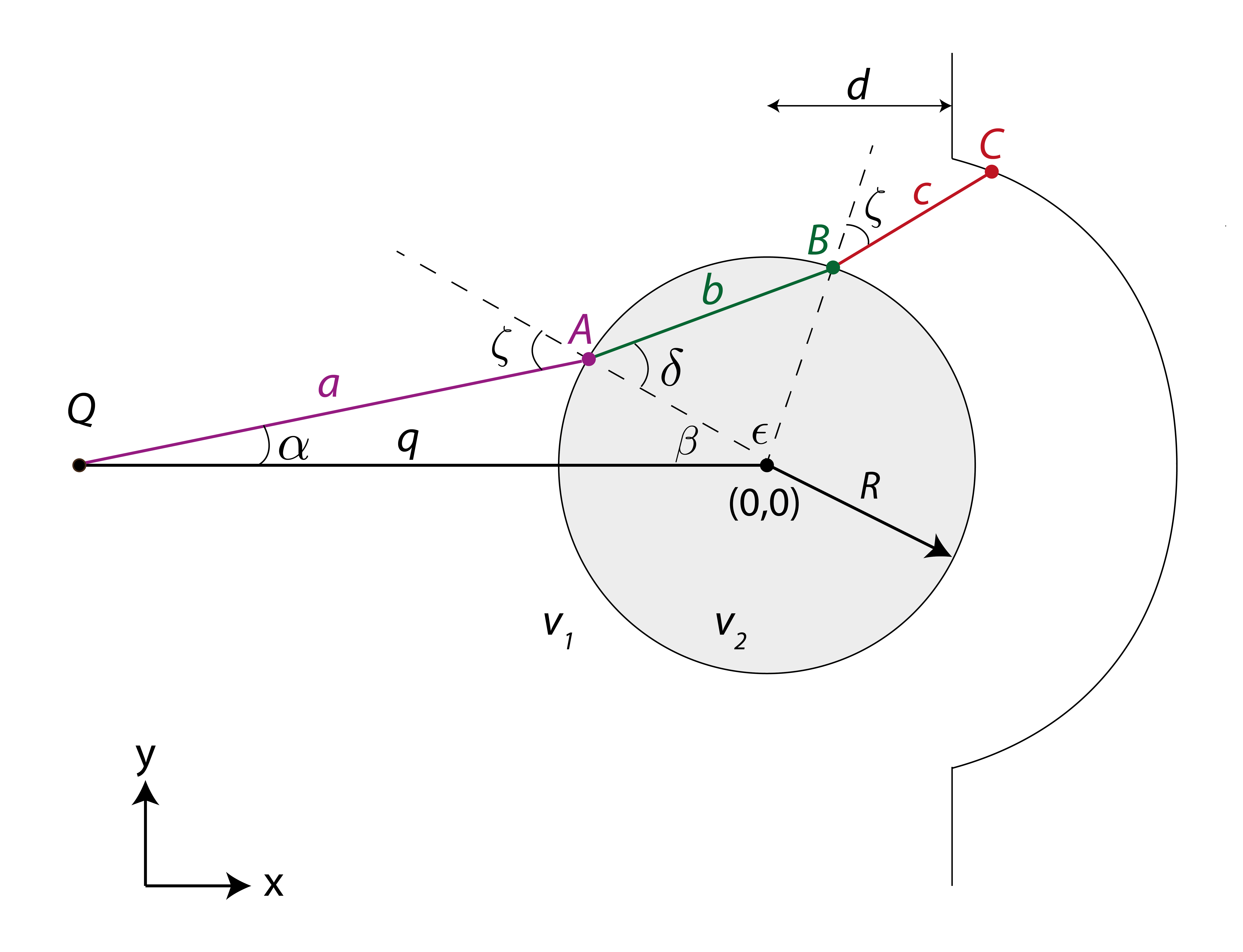}
\caption{A sketch of a ray, emitted from a point source $Q$ at $(-q,0)$, encountering a hotspot at $(0,0)$ with radius $R$. Propagation speed inside the hotspot is $v_2$ and larger than the speed outside the hotspot ($v_2>v_1$). At points $A$ and $B$ at the hotspot perimeter, the ray is refracted according to Snell's law, both when entering (point $A$) and exiting (point $B$). The black line, positioned at the right, represents the perturbed front after passing the hotspot which includes point $C$.}
\label{fig:figS1}
\end{figure}

The planar front encountering the hotspot can be regarded as originating from a point that is positioned infinitely far away. We will first consider a point source at a finite distance and then take the limit to infinity. Consider the hotspot with radius $R$ to be located at the origin $(0,0)$, and the point source $Q$ to be positioned at $(-q,0)$. One path of least time, connecting the front to point $Q$ is sketched in Fig. \ref{fig:figS1}. In analogy to geometrical optics, we consider this path as a single ray, emitted under an angle $\alpha$ from the point source $Q$. Refraction occurs both when this ray enters and when it leaves the hotspot. The ray consists of the following three parts: (1) The ray travels a distance $a$ from the source $Q$ to point $A$, located at the perimeter of the hotspot, with speed $v_1$. (2) Following refraction at point $A$, the ray travels a distance $b$ inside the hotspot, with speed $v_2$, until it reaches point $B$. (3) At point $B$, the ray is refracted again and travels a distance $c$ with speed $v_1$ outside the hotspot. At time $t=t_\textrm{total}$, it reaches the front at point $C$. Note that this description assumes $\left| \alpha \right| < \arcsin(R/q)$, for $\left|\alpha\right| > \arcsin(R/q)$ the ray originating in $Q$ does not encounter the hotspot.

The coordinates of point $A$ are given by
\begin{align}
A_x &= -q+ a \cos(\alpha)\, ,\nonumber \\
A_y &=  a\sin(\alpha).
\label{eq:A}
\end{align}
The length of the segment between $Q$ and $A$ is given by $a = q \cos(\alpha) - \sqrt{R^2 - (q \sin(\alpha))^2}$ as can be seen by considering this line segment as part of the cathetus of a right triangle with hypothenuse from $Q$ to the origin.

To obtain the coordinates of point $B$, we first compute the angle of incidence $\zeta$. From Fig. \ref{fig:figS1} we get $\zeta=\alpha+\beta$, with $\beta = \arcsin( a\sin(\alpha)/R )$. With Snell's law (equation \eqref{eq:Snell}), the angle of refraction $\delta$ is obtained as $\delta = \arcsin( v_2 / v_1 \cdot \sin(\zeta) )$. For symmetry reasons $\epsilon = \pi - 2\delta$. With $\beta$ and $\epsilon$ known, the coordinates of point $B$ can be computed as
\begin{align}
B_x &= - R \cos(\beta + \epsilon), \nonumber \\
B_y &= R \sin(\beta + \epsilon).
\label{eq:B}
\end{align}
The length of the line segment from $B$ to $C$ is given as $b=\sqrt{(B_x-A_x)^2+(B_y-A_y)^2}$.

The final point $C$ is located at the front and reached at time $t_\textrm{total}$. Its coordinates depend on $t_\textrm{total}$, which determines the length $c$ of the third line segment, from $B$ to $C$, as $t_\textrm{total}=a/v_1+b/v_2+c/v_1$. It is convenient to use the distance $d$ traveled by the planar part of the front as parameter for how far the front has propagated, instead of $t_\textrm{total}$. At time $t_\textrm{total}$, the planar part of the front (not having encountered the hotspot) has traveled a distance $q+d=v_1\,t_\textrm{total}$. Thus, $c=q+d-a-b v_1/v_2$. The angle of refraction at point $B$ is identical to the angle of incidence, $\zeta$ at point $A$. We obtain the coordinates for point $C$ as:
\begin{align}
C_x &= B_x - c \cos(\beta + \epsilon + \zeta), \nonumber \\
C_y &= B_y + c \sin(\beta + \epsilon + \zeta ).
\label{eq:C}
\end{align}
In summary, the positions of points $A$, $B$, and, importantly, the point $C$, forming part of the front, can be expressed analytically as a function of the distance of the source to the hotspot, $q$, and the angle $\alpha$ given the speeds $v_1$ and $v_2$. These results are easily modified to capture the case that the front is still inside the hotspot.

In a second step, we are taking the limit $q\rightarrow\infty$, representing a point source for a radial wave at infinity, corresponding to a planar wave encountering the hotspot. For $q\rightarrow \infty$ the maximum $\left|\alpha\right|$ approaches $0$ because of $\left| \alpha \right| < \arcsin(R/q)$. It is therefore useful to replace $\alpha$ by a parameter $x$ to parametrise the opening angle using $\alpha = x\arcsin(R/q)$. In principle, the limit $q\rightarrow \infty$ can be taken for the expressions above with $x$ being finite and parametrising the position along the initial front. Limits can also be taken in a consecutive manner resulting in:
\begin{align}
\lim_{q\rightarrow\infty}\beta &= \arcsin(x),\nonumber \\
\lim_{q\rightarrow\infty}\zeta &= \arcsin(x),\nonumber \\
\lim_{q\rightarrow\infty}\delta &= \arcsin(v_2/v_1 \cdot x),\nonumber \\
\lim_{q\rightarrow\infty}\epsilon &= \pi - 2 \arcsin(v_2/v_1 \cdot x).
\label{eq:angles}
\end{align}
Expressions for $b$, $c$ and points $A$, $B$, $C$ can be obtained straightforwardly. For point $C$ we obtain
\begin{eqnarray}
\lim_{q\rightarrow\infty}C_x & = & R \cos \left(\arcsin(x)-2\arcsin\left(\frac{v_2 x}{v_1}\right)\right) \\
& & + \left(d+R\sqrt{1-x^2}-\frac{2 v_1 R\sqrt{1-\frac{v_2^2 x^2}{v_1^2}}}{v_2}\right)\cos\left(2\arcsin(x)-2\arcsin\left(\frac{v_2 x}{v_1}\right)\right),\nonumber \\
\lim_{q\rightarrow\infty}C_y & = & -R \sin \left(\arcsin(x)-2\arcsin\left(\frac{v_2 x}{v_1}\right)\right) \nonumber \\
& & - \left(d+R\sqrt{1-x^2}-\frac{2 v_1 R\sqrt{1-\frac{v_2^2 x^2}{v_1^2}}}{v_2}\right)\sin\left(2\arcsin(x)-2\arcsin\left(\frac{v_2 x}{v_1}\right)\right). \nonumber
\end{eqnarray}

From the definition of $x$ and the condition $\left| \alpha \right| < \arcsin(R/q)$ follows $\left|x\right|<1$. However, the requirement that $\delta<\pi/2$ (``total reflection'') restricts $\left|x \right|$ further to $\left|x \right|<v_1/v_2$. Last, but not least, only those points at the front just computed are of interest that are ahead of the planar front, i.e., $C_x > d$ which restricts $\left|x \right|$ even further. Unfortunately, this inequality cannot be solved analytically. However, we observe heuristically, that the further out, the smaller the range of $x$ that contribute to the real front. In other words, for large $d$, only small $\left|x\right|$ are relevant.

\begin{figure}[h]
\includegraphics[width=0.7\textwidth]{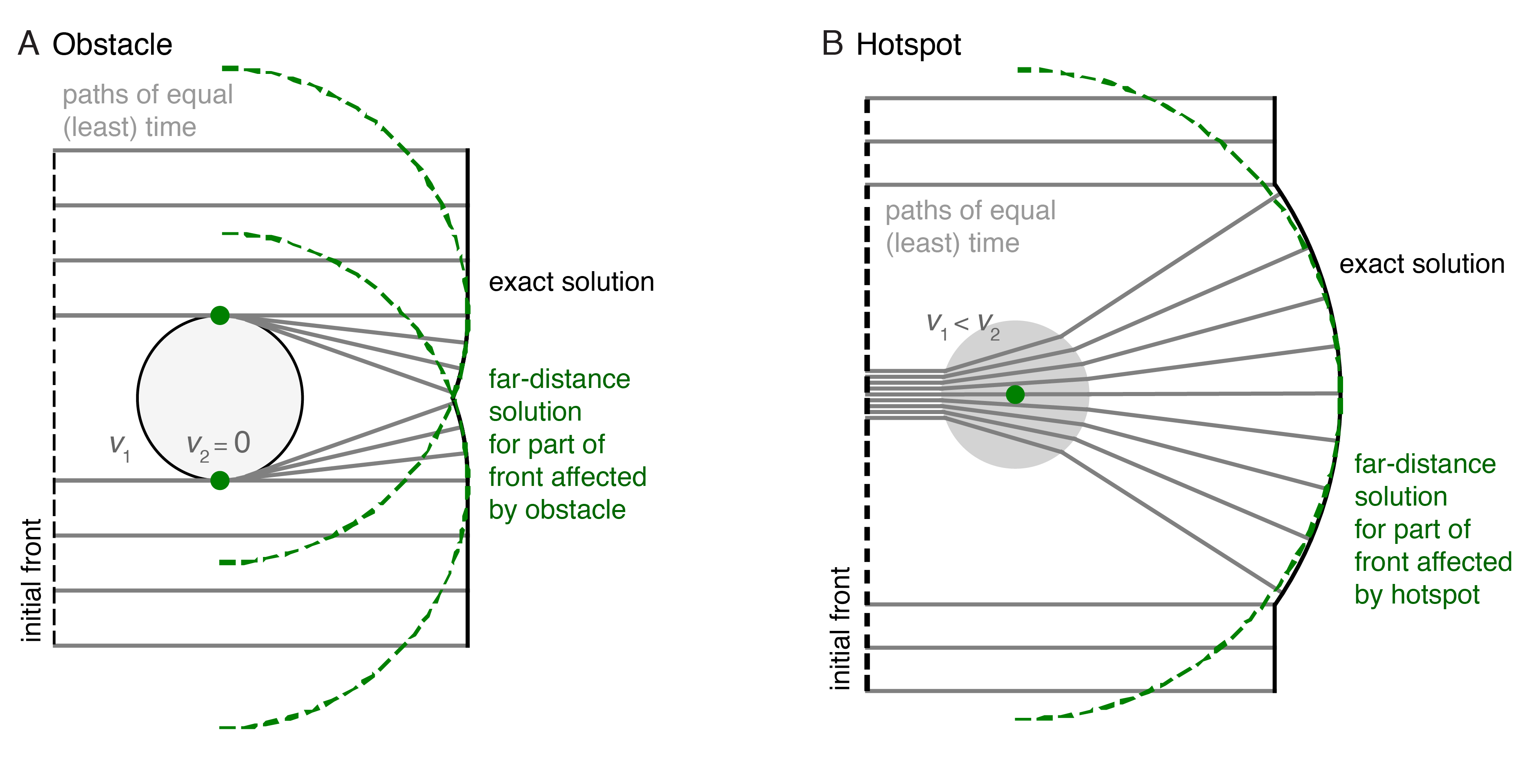}
\caption{Least-time consideration for \textbf{(A)} an obstacle and \textbf{(B)} a hotspot. The black line indicates the exact solution for the front provided in Ref.~\cite{MoebiusPLOSCB15} for the case of an obstacle and in Appendix S2 for a hotspot. The grey lines represent paths of virtual markers traveling from left to right in the same amount of time. The green dashed line indicates the front far downstream, two circular fronts originating from the sides of the obstacle and one circular wave originating from the centre of the hotspot, respectively. Please see main text for more details.}
\label{fig:figS2}
\end{figure}

\begin{figure}
\includegraphics[width=0.52\textwidth]{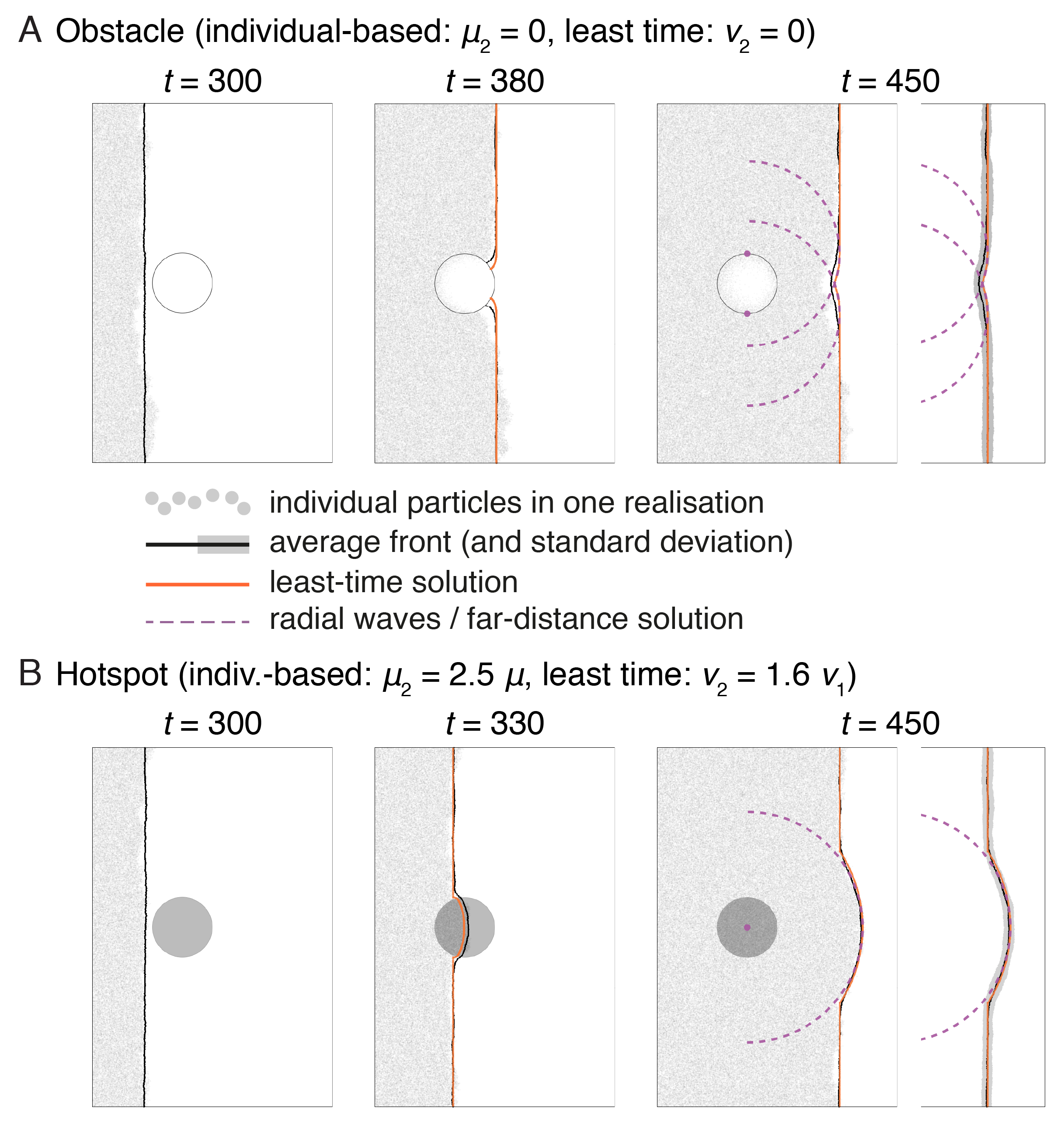}
\caption{Effects of spatially varying birth rate instead of diffusion coefficient (illustrated in Fig. 2C,D of the main text.) \textbf{(A)} Results of the individual-based simulation with an obstacle (white circle) with radius $R = 50$ and within which birth rate is set to $\mu=0$ (grey dots), overlaid by the average front obtained from multiple realisations (black line, outside the obstacle), the least-time solution (orange line), and the far-distance solution (radial waves, purple dashed lines) Ref.~\cite{MoebiusPLOSCB15}). Right-most panel indicates standard deviation to average front instead of individual particles. \textbf{(D)} Similar to panel \textbf{(B)}, but the obstacle is replaced by a hotspot (grey circle) with radius $R=50$ and a birth rate $2.5$ times larger than outside ($\mu_2=2.5 \mu$).}
\label{fig:figS3}
\end{figure}

\begin{figure}
\vspace{1.0cm}
\includegraphics{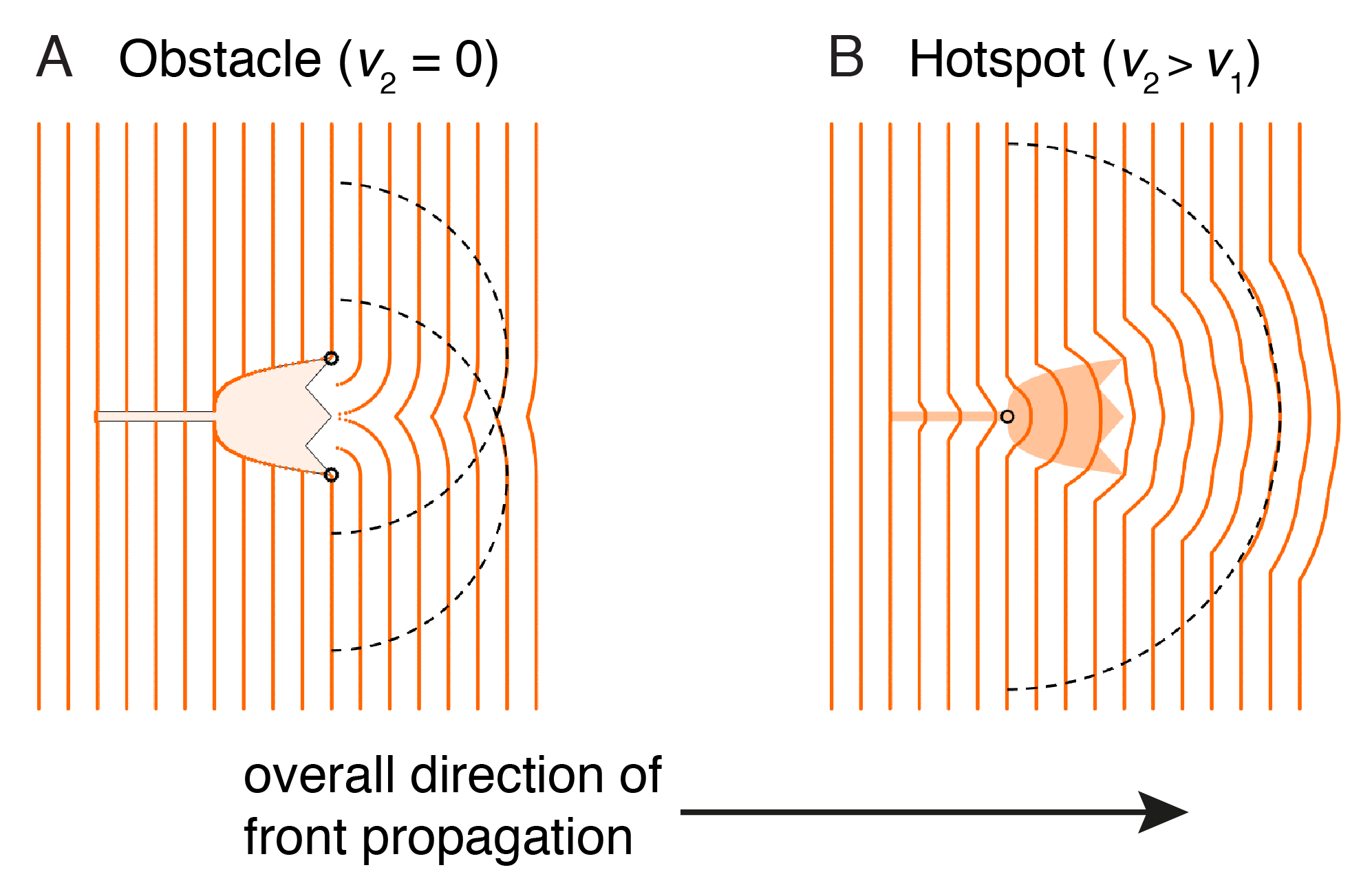}
\caption{\textbf{(A)} Front shape at different positions relative to a tulip-shaped obstacle. $v_1$ is the background speed and $v_2=0$ is the speed inside the obstacle. The dashed lines represent half-circles originating from the sides of the tulip at its widest point. \textbf{(B)} Like panel (A), but for a tulip-shaped hotspot, where $v_2 = 1.2\, v_1$. The dashed line indicates a half-circle originating at the half-length of the hotspot with radius given by Eq.~4 of the main text.}
\label{fig:figS4}
\end{figure}

For $x=0$ we obtain $C_x=d+2R(1-v_1/v_2)$ and $C_y=0$ as expected. We find that a circle whose centre coincides with the hotspots centre and with radius $d+2R(1-v_1/v_2)$ represents a good heuristic solution for the far-field limit (see Fig. 2, Fig. 3B, and Eq. 4 in the main text as well as Figs. \ref{fig:figS2}-\ref{fig:figS4}).

\subsection{Shortest path in presence of multiple consecutive rods}

We compute the front speed in a configuration of multiple rods which are to be seen as elliptical obstacles with infinite aspect ratio as illustrated in Fig. 5C in the main text. The slow-down of the front is determined by the increase in path length, relative to the straight path. We here compute the slow-down expected if the path takes course along the corners of all consecutive rods. However, shorter paths may exist as illustrated in Fig. 5C of the main text. The slow-down computed when assuming passing all consecutive rods should therefore represent a lower limit for the actual front speed in the presence of many rods.

We first compute the slow-down from one rod to another and then average over all possible configurations to capture the effect of a very large number of randomly oriented rods. Let us consider a path that originates at the right side of a given rod of width $b$, as in Fig. 5C in the main text. The path could pass the right or left corner of the following rod, depending on how much the projections of the two rods of width $b$ overlap. If this overlap is smaller than $b/2$, the shortest path grazes the left corner, otherwise it grazes the right corner.

We are now interested in the probability of encountering the next rod at a distance $x$ away with overlap $y$. First, we recognise that the rods are randomly distributed in the direction of overall front propagation. The density of rods on a straight line is given by $\rho \cdot b$, with $\rho$ the number density of the rods. As a result, the probability of encountering the next rod at a distance $x$ is exponentially distributed as $q(x) = \rho b \exp \left( -\rho b x \right)$. The probability that the overlap is $y$ is given by $p(y) = \frac{2}{b}$. Here, we restricted $y$ to be smaller than $b/2$, but included a factor $2$ to take into account that the path can encounter either a left or a right corner as discussed above. We can now compute the average path length between two consecutive rods as a function of $b$ and $\rho$:
\begin{equation}
t(\rho,b) = \int_0^{b/2} \textrm{d}y \int_0^\infty \textrm{d}x\cdot p(y)\cdot q(x)\cdot \sqrt{y^2+x^2}. \label{eq:tav}
\end{equation}
We can compute the path length in the direction parallel to front propagation simply by replacing $\sqrt{y^2+x^2}$ in equation \eqref{eq:tav} by $x$. The relative front speed is given by the ratio between this parallel path length and $t(\rho,b)$. The final expression for the relative front speed $\nu$ depends only on the dimensionless quantity $\omega = \rho b^2$:
\begin{equation}
\nu(\omega)=\frac{1}{2 \omega^2\int_0^{1/2}\textrm{d}y' \int_0^\infty \textrm{d}x' \exp(-\omega x')\sqrt{x'^2+y'^2}}.
\end{equation}

\subsection{1D succession of patches}

\subsubsection{A travelling wave is supported and established in all patches.} 

Consider a front that travels either with speed $v_1$ in the background environment or with speed $v_2$ inside the patches, which occupy a fraction $\phi$ of the environment; see Fig. 1B of the main text. Let $L$ be a distance large enough to incorporate a large number of patches. To travel that distance, the time $L/v_\textrm{eff}=L(1-\phi)/v_1+L\phi/v_2$ is needed. Thus, the effective speed is given by
\begin{equation}
v_\textrm{eff}=\frac{1}{(1-\phi)/v_1+\phi/v_2}\, ,
\end{equation}
which is the (weighted) harmonic mean of the two front speeds. Note that the effective speed is independent of the size of the patches.
The relative speed $\nu_h$ discussed in the main text follows from dividing by $v_1$ and using $v_2=\gamma v_1$:
\begin{equation}
\nu_h=\frac{1}{\phi/\gamma+(1-\phi)}\, .
\end{equation}

\subsubsection{A travelling wave is established in favourable regions only.}

Consider a traveling population wave of speed $v_1$. Inside obstacles of size $d$, the population wave cannot be sustained, but individuals can diffuse. These obstacles occupy a fraction $\phi$ of the environment. The time to travel across a large distance $L$ that includes $N_\textrm{obs}$ large obstacles can be estimated as
\begin{equation}
\frac{L}{v_{eff}}\sim \frac{L(1-\phi)}{v_1}+\frac{d^2}{D}N_\textrm{obs}\,.
\end{equation}
Note that we consider scales large enough to neglect the times to establish the traveling population wave.

Because $N_\textrm{obs}d=L\phi$:
\begin{equation}
v_{eff} \sim \frac{1}{(1-\phi)/v_1+d\phi/D}\,.
\end{equation}
The front can be `arbitrarily' slowed down by increasing $d$. In contrast to the case where both types of environments support a travelling wave, the size of the obstacles does matter here -- but not the size of the complement, the favourable patches.

\newpage
\section{Supplementary Figures}
\begin{figure}
\vspace{1.0cm}
\includegraphics{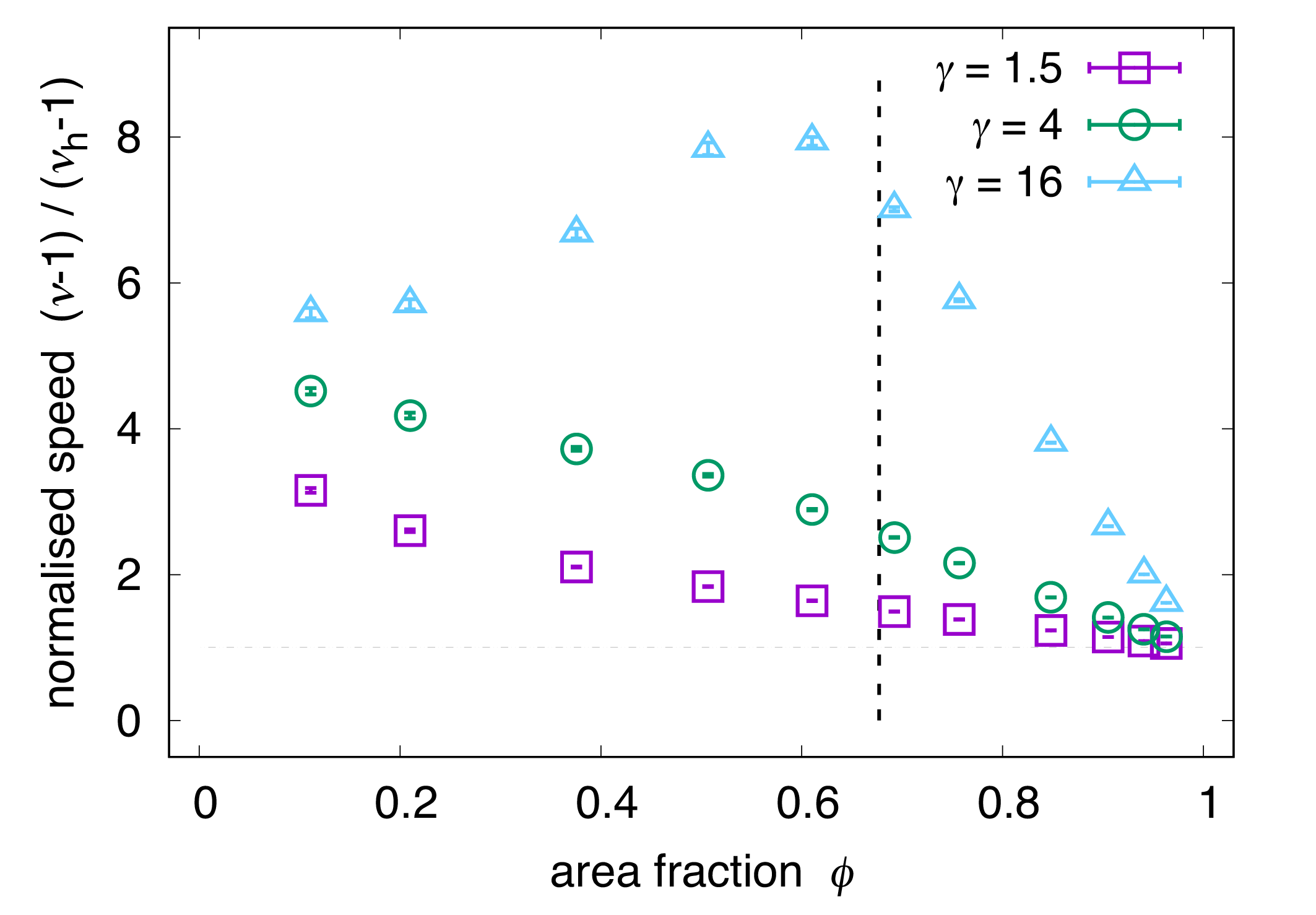}
\caption{Normalised front speed $(\nu-1)/(\nu_h-1)$ as function of hotspot strength $\gamma$ and area fraction $\phi$ where $\nu_h=\left(\phi/\gamma+(1-\phi)\right)^{-1}$ is the weighted harmonic mean of front speeds inside and outside of hotspots. Values larger than $1$ indicates speed-up attributed to the two-dimensionality of the system. The dashed vertical line corresponds to the percolation threshold in an infinite system ($\phi=0.68$).}
\label{fig:figS5}
\end{figure}
\begin{figure}
\vspace{1.0cm}
\includegraphics{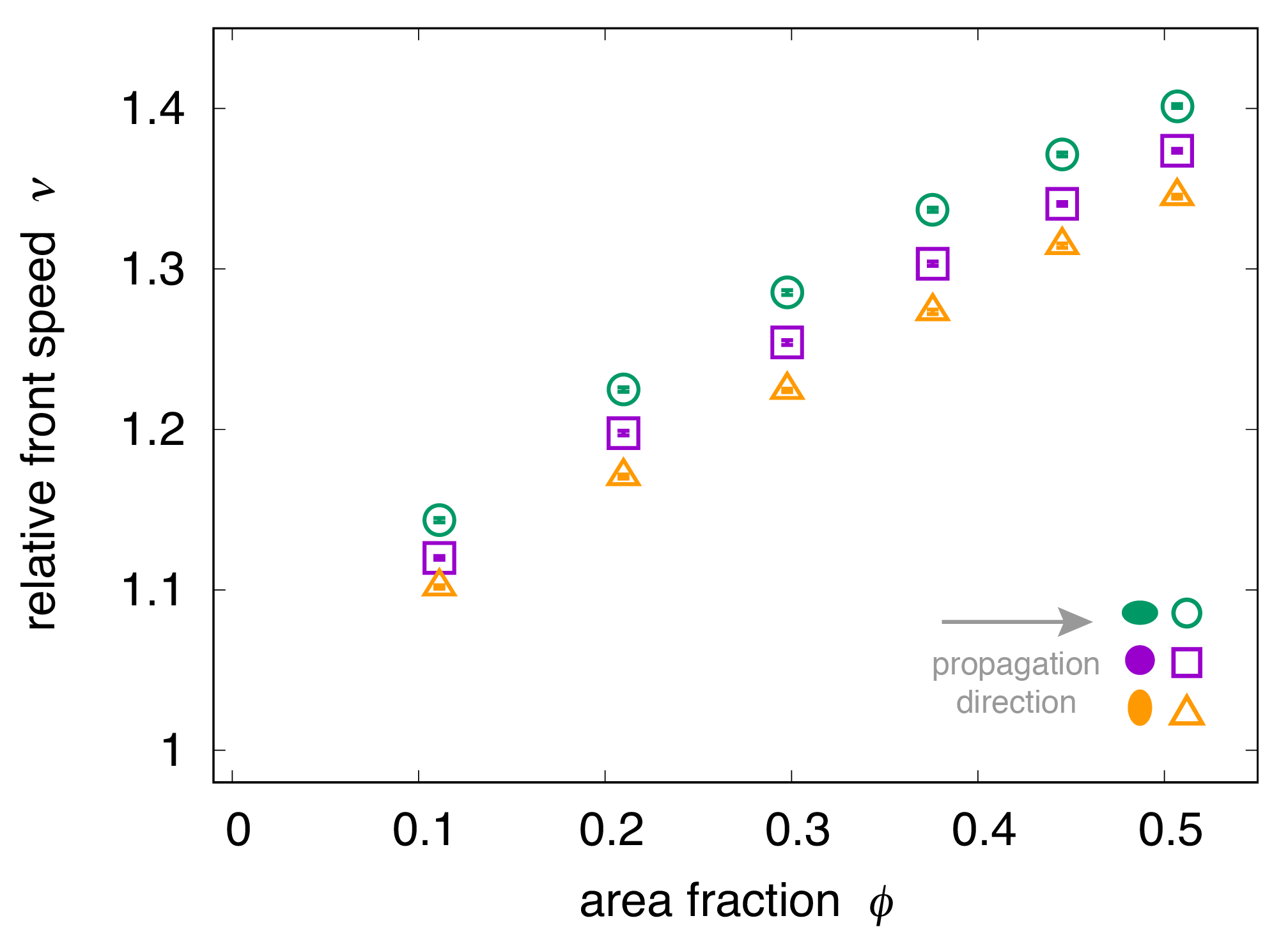}
\caption{Relative front speed $\nu$ as a function of area fraction $\phi$ for weak elliptical hotspots with strength $\gamma=1.5$ for three different aspect ratios (purple: $1$, green: $3/2$, orange: $2/3$). See Fig.~6C of the main text for an equivalent plot, but for strong hotspots with $\gamma=16$.}
\label{fig:figS6}
\end{figure}
\begin{figure}
\vspace{1.0cm}
\includegraphics{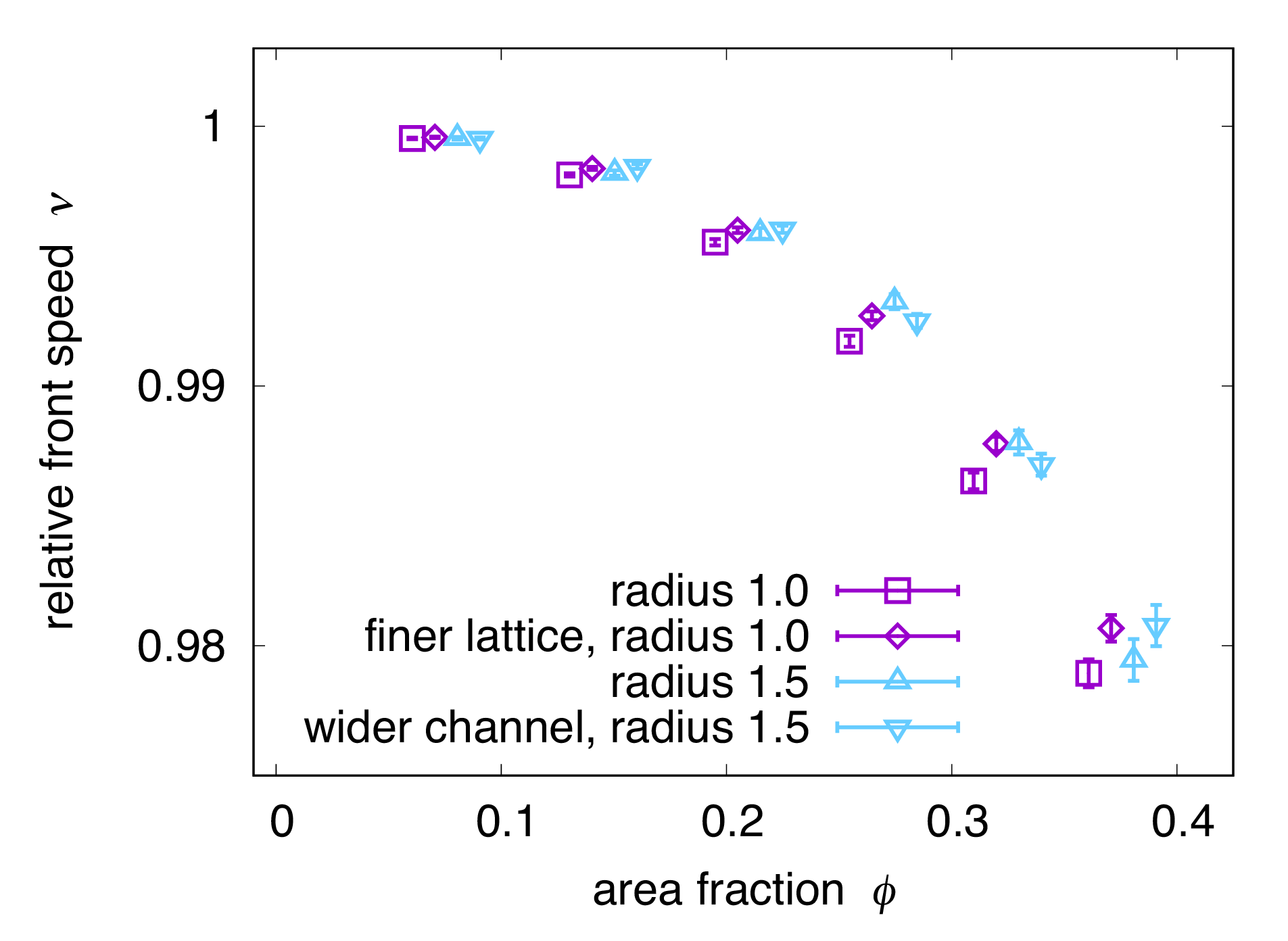}
\caption{Effect of changing lattice constant and channel width on numerically determined front speed in the presence of obstacles, compare to Fig.~5A of the main text. Relative front speed $\nu$ obtained by numerically solving the Eikonal equation as a function of $\phi$, the area fraction covered by obstacles. For small obstacles, the effect of lattice constant was investigated (default: $1/15$, finer: $1/22$), for large obstacles the effect of channel width (default: $50$, wider: $75$). Data points are offset slightly along the abscissa for clarity.}
\label{fig:figS7}
\end{figure}
\begin{figure}
\vspace{1.0cm}
\includegraphics{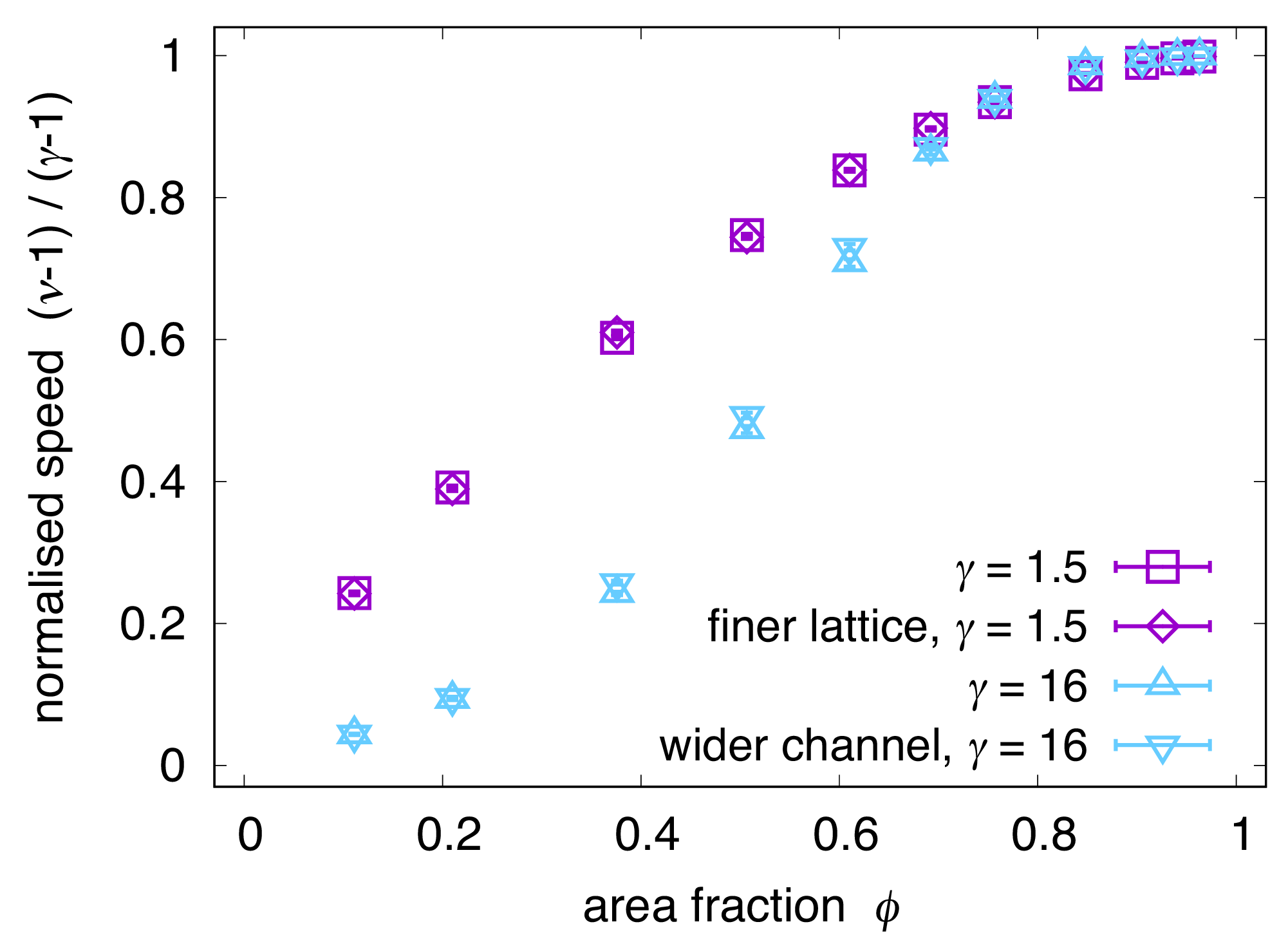}
\caption{Effect of changing lattice constant and channel width on numerically determined front speed in the presence of hotspots, compare to Fig.~6A of the main text. Relative front speed $\nu$ obtained by numerically solving the Eikonal equation as a function of $\phi$, the area fraction covered by hotspots. For weak hotspots, the effect of lattice constant was investigated (default: $1/15$, finer: $1/22$), for strong hotspots the effect of channel width (default: $50$, wider: $75$).}
\label{fig:figS8}
\end{figure}

\FloatBarrier
\section{Supplementary Videos}

\begin{video}
\caption{Time lapse of the individual-based simulation with an obstacle (white circle) with radius $R = 50$ and $D_2=0$ (grey dots), overlaid by the average front obtained from multiple realisations (black line, outside the obstacle), the least-time solution (orange line), and the far-distance solution (radial waves, purple dashed lines). For individual snapshots see Fig.~2C of main text.}
\end{video}

\begin{video}
\caption{Time lapse of the individual-based simulation with an hotspot (grey circle) with radius $R = 50$ and $D_2=2.5 D$ (grey dots), overlaid by the average front obtained from multiple realisations (black line, outside the obstacle), the least-time solution (orange line), and the far-distance solution (radial waves, purple dashed lines). For individual snapshots see Fig.~2D of main text.}
\end{video}

\begin{video}
\caption{Front propagation through a system of randomly-placed wide obstacles (major axis parallel to front, aspect ratio $2/3$) with area fraction $\phi\approx 0.34$. %
The front is a constructed using the least-time approach, where the Eikonal equation is solved numerically.}
\end{video}

\begin{video}
\caption{Front propagation through a system of randomly-placed long obstacles (minor axis parallel to front, aspect ratio $3/2$) with area fraction $\phi\approx 0.51$. %
The front is a constructed using the least-time approach, where the Eikonal equation is solved numerically.}
\end{video}

\begin{video}
\caption{Front propagation through a dilute system $\phi\approx 0.21$ %
of randomly placed hotspots of strength $\gamma=16$. The front is a constructed using the least-time approach, where the Eikonal equation is solved numerically. In this dilute regime, hotspots accelerate the front locally.}
\end{video}

\begin{video}
\caption{Front propagation through a dense system $\phi\approx 0.85$ %
of randomly placed hotspots of strength $\gamma=16$. The front is a constructed using the least-time approach, where the Eikonal equation is solved numerically. In this dense regime, the front propagates through a network of hotspots.}
\end{video}

\begin{video}
\caption{Front propagating in and transitioning from a dense to a dilute hotspot configuration, simulated using the least-time approach. The interface between the dense and dilute domain is tilted by 45\degree. When the front passes this interface, the difference in effective front speed leads to `refraction' of the front towards the dilute configuration. See caption of Fig. 7 for more details.}
\end{video}

\FloatBarrier
\bibliography{bib_obstacles_hotspots}